\documentstyle[aps,pre,graphicx,amsmath,multicol]{revtex}

\begin{document}
\draft
\voffset=2cm

\newcommand{\btau}{\mbox{\boldmath{$\tau$}}}


\title{Granular flow down an inclined plane: Bagnold scaling and
rheology}

\author{Leonardo E. Silbert$^1$, Deniz Erta{\c s}$^2$, Gary
  S.~Grest$^1$, Thomas C.~Halsey$^2$, Dov Levine$^3$, and Steven J.
  Plimpton$^1$}

\address{$^1$ Sandia National Laboratories, Albuquerque, New Mexico
  87185\\ $^2$ Corporate Strategic Research, ExxonMobil Research and
  Engineering, Annandale, New Jersey 08801\\$^3$ Department of
  Physics, Technion, Haifa, 32000 Israel}

\date{\today}

\maketitle

\begin{abstract}
  We have performed a systematic, large-scale simulation study of
  granular media in two- and three-dimensions, investigating the
  rheology of cohesionless granular particles in inclined plane
  geometries, i.e., chute flows. We find that over a wide range of
  parameter space of interaction coefficients and inclination angles,
  a {\it steady state} flow regime exists in which the energy input
  from gravity balances that dissipated from friction and inelastic
  collisions. In this regime,
  the bulk packing fraction (away from the top free surface and the
  bottom plate boundary) remains constant as a function of depth $z$,
  of the pile. The velocity profile in the direction of flow
  $v_{x}(z)$ scales with height of the pile $H$, according to
  $v_{x}(z)\propto H^{\alpha}$, with $\alpha=1.52\pm0.05$. However,
  the behavior of the normal stresses indicates that existing simple
  theories of granular flow do not capture all of the features
  evidenced in the simulations.
\end{abstract}
\pacs{46.55.+d, 45.70.Cc, 46.25.-y}


\begin{multicols}{2}

\section{Introduction}
It is tempting to regard the behavior of granular materials as being a
problem in engineering or applied science,
inasmuch as the fundamental laws governing their
constituent particles are well known. Being comprised of
macroscopically large grains, granular materials obey classical
mechanics, although the existence of friction and inelastic collisions
complicates matters.  However, while it is true that the
collision of two grains is analytically tractable, an aggregate of
such grains is a many-body system, whose macroscopic behavior
cannot be simply related to the laws controlling individual constituents.  

For this reason, a continuum treatment is often adopted, in which the
variables are averaged properties whose governing equations are
derivable, in principle, from the known microscopic laws.  Among these
averaged variables are the density $\rho$ and the stresses
$\sigma_{\alpha\beta}$, which obey the Cauchy equations that
enforce momentum conservation (or force balance if there are no
accelerations). However, this set of equations is insufficient
to solve for the stresses, since there are too few equations: in $D$
dimensions, there are $D(D+1)/2$ independent 
components of $\sigma_{\alpha\beta}$
(which is a symmetric tensor), but only $D$ equations of momentum
conservation.  Therefore, the Cauchy equations must be augmented by
additional constitutive relations, possibly history-dependent,
which tell how the material in
question responds to the application of a force.  It is in these
constitutive relations that the specifics of the material in question
come into play.  In the case of steady state flow, which we will
consider in this paper, constitutive equations would relate the
strain-rate, $\dot\gamma_{\alpha\beta}$, to the stress.

In 1954, Bagnold \cite{bagnold1} proposed that in inertial granular
flow, the shear stress is proportional to the square of the
strain-rate:
\begin{equation}
\label{eqbagnold}
\sigma \propto \dot\gamma^2.
\end{equation}
His argument, applied to the case of bulk granular flow, is
predicated on a constant density profile.  In practice, the presence
of significant finite-size or wall effects often obscures Bagnold
scaling.  In this study, we report a set of numerical simulations of
bulk granular flow down an inclined plane, so-called ``chute flow'', in
two and three dimensions.  The geometry is simple: a layer of bulk
granular material is placed on a flat plane of area $A$ 
(or line of length $L$ in 2D) on which grains have been glued, 
so as to form a rough base. The thickness of the layer is measured
in terms of the pile height parameter $H\equiv N d^2/A$ (or 
$N d/L$ in 2D), where $N$ and $d$ are the total number of particles 
and their diameter, 
respectively. The plane is inclined at an angle $\theta$ and the 
flow is observed.  The parameters controlling the flow are the 
macroscopic variables $\theta$ and $H$, as well as the microscopic 
variables determining the nature of interaction between two grains, 
such as grain friction $\mu$ and coefficient of restitution $\epsilon$.

In Ref.~\cite{deniz2}, we provided a summary of our simulations in two
and three dimensions; in this paper we expand on these results both in
depth and breadth for the case of steady state flow.  The results
obtained reveal the rich and surprising nature of the collective
behavior of the system. For certain values of the parameters, we
observe Bagnold scaling in stable steady state flow, with a constant
density profile independent of depth.  However, we also saw surprising
examples of self-organization, including the flow-induced
crystallization of a disordered state into one with much lower
dissipation. In this regime (systems flowing on moderately smooth
bottom surfaces), we found reentrant disordering as well, and even
oscillations between ordered and disordered states.  The effects of
bottom surfaces are thoroughly discussed in a separate work
\cite{leo8}. In this paper, we concentrate on rough bottom surfaces
for which the behavior is simpler.

These simulations also allow us to investigate more subtle aspects of
chute flow, such as hysteresis in the angle of repose and normal
stress inequalities not accounted for by any conventional continuum
theory. Additionally, we were able to look for surface and bulk
instabilities to flow at the angle of repose. In particular, we find
that although the Bagnold rheology of flow near the angle of repose is
a bulk rheology, the fundamental instability inducing the flow in
three dimensions appears to be an instability of the surface layers of
the granular medium.

Because granular materials are so common in nature, existing on many
different length scales, there is a great amount of experimental data
on a wide range of dynamic situations; shear flow and vibration
experiments \cite{hsiau2,behringer2,choplin1,nagel3}, and studies of
geological debris flows \cite{takahashi1}, to name just a few
\cite{campbell5,jenkins1}. There have been several moderately
well-characterized experimental studies of granular flow down an
inclined plane under laboratory conditions
\cite{durian1,pouliquen2,hungr1,drake1,azanza1,walton3}.  Yet for all
the intense activity in this field over the years, the rheology of
granular systems still remains a largely unsolved problem.

There has been some work on continuum modeling of chute flow; for a
review of continuum based ideas see Savage \cite{savage2} and
references therein. Other theoretical analyses (sometimes combined
with case-specific simulation verification), specifically applied to
chute flow geometries, attempt to calculate density and velocity
profiles \cite{brennen1,mills4,khakhar1,rapaport1}, but a general
consensus on the qualitative features of these profiles has yet to be
reached.  

The state of the art of computer simulations of chute flow is much
less satisfactory because of the enormous equilibration times needed
to set up steady flow. In three dimensions, simulations have been
performed for rather thin piles, which provides insight into only a
small region of phase space \cite{walton3,walton1,walton2}.
Simulations of two-dimensional flows also report on small systems, and
it is unclear whether these studies are carried out in the steady
state regime or whether the data reported are transient
\cite{poschel1,zheng1,zheng2,wolf2}. The
basics behind granular simulations are available in Refs.
\cite{cundall1} for 2D and \cite{walton2} for 3D. 

Our simulations attempt a systematic 3D study of chute flows.
Unfortunately, we probe regions of phase space difficult to access in
experiment.  In a typical 3D experiment the flow is induced through a
hopper-feeder mechanism which controls the flow rate of the system,
but not the thickness of the flowing sample, which is chosen
spontaneously by the system.  Thus, much experimental data is for
flowing piles 10-15 particles high, whereas most of our simulations
focus on moderate to thick piles, greater than 30 particles.
Simulation results for systems smaller than 10 - 15 particles high do
not show the same scaling as that for thicker systems \cite{durian1}.
Also, 3D experiments are usually carried out in narrow channels of
order 10 particles wide or less, where side-wall effects may have
a significant role in the observed behavior. Our simulations are
periodic in this, the vorticity, direction, and we have yet to study
wall effects. We suspect that most discrepancies that may exist
between different experimental and simulation studies are due to such
differences in the detailed nature of the systems studied.

\begin{figure}
\begin{center}
\includegraphics[width=2.9in]{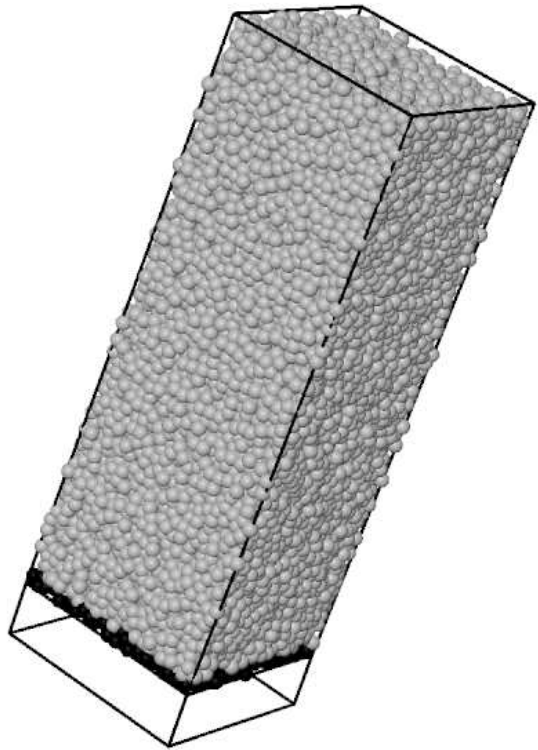}
\caption{\em Typical 3D snapshot for chute flow: N=24,000, with 
bottom surface dimensions 20{\rm x}20 diameters shown by the black particles fixed
to the bottom plate; tilt angle $\theta=24^{o}$, 
coefficient of restitution $\epsilon=0.88$, and friction coefficient
$\mu=0.50$. Flow is directed down the incline.}
\label{config_3d} 
\end{center}
\end{figure}

Because of the complexity of flowing granular systems, it is useful to
first define the region of study. In order to determine the phase
boundaries of fully developed, steady state flow, we have performed a
series of simulations of inclined plane gravity driven flows in two
and three dimensions in an attempt to define the region of phase space
for which steady state flows exist. A typical configuration snapshot
in 3D, defining the computational geometry, is shown in
Fig.~\ref{config_3d}. 

In our simulations, initiation of flow is achieved by tilting at a
large angle ($24-30^{o}$) to induce flow.  This procedure removes any
configuration construction history effects.  We then reduce the
inclination to a lower angle and allow the simulation to run until we
observe a steady state flow regime (if one exists). We define steady
state as flow wherein the energy input from gravity balances that
dissipated from friction and collisions; so that the total kinetic 
energy of the system reaches a macroscopically constant value.
In this case, the results are independent of sample history. 

\begin{figure}
\begin{center}
\includegraphics[width=2.9in]{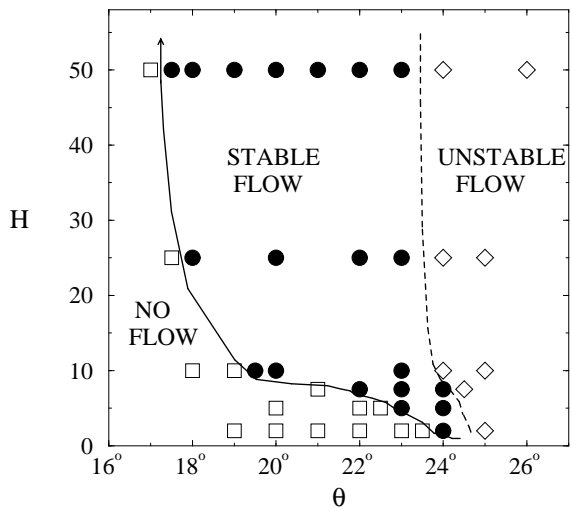}\hfil
\includegraphics[width=2.9in]{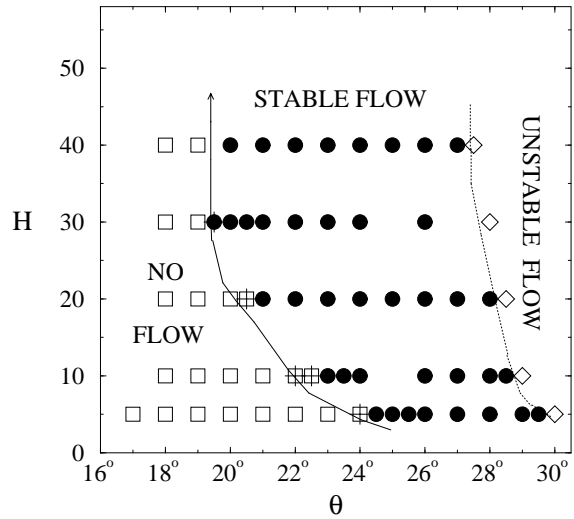}\\
\hfil(a)\hfil\hfil(b)\hfil\\
\bigskip
\caption{\em Phase behavior of granular particles 
  in chute flow geometry, characterized by pile height $H$ vs. tilt
  angle $\theta$ for monodisperse systems in (a) 2D with $\mu=0.50$
  and $\epsilon=0.92$ (identified as Model L2 in Table \ref{table1}),
  and (b) 3D with $\mu=0.50$ and $\epsilon=0.88$ (identified as Model
  L3). Both figures are for the spring dash-pot interaction model with
  rough bottom surface. Solid circles indicate region of steady state
  flow, open symbols correspond to no flow or unstable flow. In 3D we
  have identified hysteretic flow as {\rm x}.}
\label{phase} 
\end{center} 
\end{figure}

In Fig.~\ref{phase}, we draw phase boundaries for both two and three
dimensional flows as a function of the external control parameters:
tilt angle $\theta$ and pile height $H$. This should be compared 
to a similar experimental determination recently obtained by 
Pouliquen \cite{pouliquen1}. The salient features are the existence
in both 2D and 3D of three principal regions, corresponding to 
{\it no flow}, {\it stable flow}, and {\it unstable flow}.
For a system of given thickness and 
fixed microscopic interaction parameters, these three regions  
are separated by two angles: $\theta_r$, the angle of repose, 
and $\theta_{max}$, the {\it maximum stability angle}, the largest
angle for which stable flow is obtained, shown by a solid 
and dashed lines in Fig.~\ref{phase}, respectively.

\end{multicols}

\begin{multicols}{2}
  
For $\theta < \theta_ r$, granular flow cannot be sustained.  In the
region $\theta_r < \theta < \theta_{max}$, we obtain steady state
flow with packing fraction independent of depth.  The region of
constant packing fraction in the flowing material for steady state
systems is accompanied by a smoothly varying, non-linear velocity
profile. For $\theta>\theta_{max}$, the development of a shear
thinning layer at the bottom of the pile results in lift-off and
unstable acceleration of the entire pile. The exact locations of
these phase boundaries depend on the model parameters such as $\mu$
and $\epsilon$. For instance, in 2D, if $\epsilon$ is reduced from
$0.92$ to $0.82$, the maximum angle of steady-state flow increases
from $\approx 23^{o}$ to $26^{o}$. Similarly, reducing $\mu$
typically reduces the range of stable flow.

It is well known that granular systems exhibit hysteresis. Such
behavior is usually attributed to system preparation and associated
history effects \cite{behringer3}. Although we observe three
distinct regimes, the behavior close to the phase boundaries is
sensitive to the procedure for the initiation of flow. Indeed, we have
observed hysteresis in our 3D simulations when approaching
$\theta_{r}$ from either side, particularly for thinner piles. The
hysteresis was significantly reduced upon increasing pile height $H$.
Crystallization of the 2D monodisperse pile upon the arrest of flow
was primarily responsible for the large hysteresis observed in that case.

Besides the phase diagram, our most important results concern the
detailed structure and rheology of the steady-state flowing regime.
In this regime, we do see a constant density profile  with height, 
as well as the Bagnold scaling of Eq.(\ref{eqbagnold}).
The amplitude of the strain rate goes to zero at the angle of
repose; thus relations such as Eq.(\ref{eqbagnold}) possess an
additional strong angular dependence. 

We also analyzed the normal stresses in the flowing state, 
and found a number of results, most notably that the normal
stress perpendicular to the free surface, $\sigma_{zz}$,
is approximately, but not exactly, equal to the normal stress 
parallel to the flow, $\sigma_{xx}$. 

There are two fundamental puzzles in these results for the rheology
of chute flow. The first, and smaller, puzzle is the appearance
of an anomalous normal stress difference $\sigma_{zz}-\sigma_{xx}$. 
We have been unable to define a simple, local, dimensionally
consistent and rotationally invariant constitutive relation
connecting $\sigma$ to $\dot\gamma^2$ that recovers this behavior.

The second, and deeper puzzle, is the relationship between the rheology
and the Coulomb yield criterion. As the angle of repose is approached
from above, the amplitude of the flow goes to zero, but the tensor 
structure of $\sigma$ remains approximately liquid-like, instead of
recovering the large normal stress difference characteristic of
the Coulomb yield criterion, which presumably applies to the static
pile at the angle of repose. The one exception to this observation is
the surface stress in three dimensions, where the normal stress
differences do become large as the angle of repose is approached,
suggesting that surface yield may control the failure of the static
state.

In the bulk, however, we are left with a transition to a static state
that appears continuous in shear rate, but is apparently discontinuous
in normal stress. We do not believe that an understanding of chute
flow rheology is possible without resolving this seeming paradox.

We present the simulation scheme in Section \ref{simmed}, detailing
the inter-particle force laws.  In Section \ref{results}, we report 
our comprehensive simulation analysis, including the behavior of 
the density and velocity profiles for our systems with varying 
interaction parameters. In Section \ref{secstress}, we present a
detailed discussion of stress analysis and
rheology of chute flow systems. In Section \ref{secconc} we 
summarize our findings.

\section{Simulation Methodology}
\label{simmed}

We use the methods of molecular dynamics to perform 2D and 3D
simulations of granular particles. For this study we model $N$
mono-disperse spheres of diameter $d$ and mass $m$, supported on the
$xy$-plane by a rough bed. The computational geometry of the present
3D system consists of a rectangular box with periodic boundary
conditions in the $x$ (flow)- and $y$ (vorticity)-directions and
constrained in the vertical $z$-direction by a fixed rough, bottom
wall and a free top surface, as in Fig.~\ref{config_3d}.
Simulations in periodic
cells attempt to study flow down infinitely long and wide chutes,
while using a finite number of particles.

In 3D, the fixed bottom is constructed from a random conformation
of spheres of the same diameter $d$ as those in the bulk by taking a
slice with areal fraction very close to random close packing
(approximately 1.2 particle diameters thick), from a previously packed
state. This simulates an experimental procedure whereby glue is
spread over the original smooth chute surface and particles are then
sprinkled onto this surface to construct a rough floor approximately
one particle layer thick. For 2D studies, the bottom wall is
constructed from a regular array of spheres of diameter $2d$ and
particle motion is restricted to the $xz$-plane.

We employed a contact force model that captures the major features of
granular interactions. In 2D, interactions between (projected) spheres
are modeled using a linear spring model with velocity-dependent
damping (the spring-dashpot interaction) and {\it static friction}. In 
3D the spring-dashpot model and static friction are also used, as well
as Hertzian contact forces with static friction.  In the presentation
of the results, we will specify which model is employed, and discuss
the differences.

The implementation of the contact forces, both the normal forces and
the shear (friction) tangential forces, is essentially a reduced
version of that employed by Walton and Braun \cite{walton1}, developed
earlier by Cundall and Strack \cite{cundall1}. More recent versions of
these models now exist \cite{zhang1,zhang2,moreau1}.  We ignore
hysteretic effects between loading or unloading normal contacts and we
do not differentiate between frictional directions at the same contact
point at different time-steps as does Walton\cite{walton2}.

Static friction is implemented by keeping track of the elastic shear
displacement throughout the lifetime of a contact. For two contacting
particles $\{i,j\}$, at positions $\{{\bf r}_{i}, {\bf r}_{j}\}$, with
velocities $\{{\bf v}_{i},{\bf v}_{j}\}$ and angular velocities 
$\{{\mathbf \omega}_{i},{\mathbf \omega}_{j}\}$, the force on 
particle $i$ is computed as follows: the normal compression 
$\delta_{ij}$, relative normal velocity ${\bf v}_{n_{ij}}$, relative
tangential velocity ${\bf v}_{t_{ij}}$ are given by

\begin{eqnarray}
\delta_{ij}&=&d-r_{ij}, \\
\label{equation1}
{\bf v}_{n_{ij}}&=&({\bf v}_{ij}\cdot{\bf n}_{ij}){\bf n}_{ij}, \\
\label{equation2}
{\bf v}_{t_{ij}}&=&{\bf v}_{ij}-{\bf v}_{n_{ij}}-\frac{1}{2}
({\mathbf \omega}_{i}+{\mathbf \omega}_{j})\wedge{\bf r}_{ij},
\label{equation3}
\end{eqnarray}

\noindent
where ${\bf r}_{ij}={\bf r}_{i}-{\bf r}_{j}$, ${\bf n}_{ij}=
{\bf r}_{ij}/r_{ij}$, with $r_{ij}=|{\bf r}_{ij}|$, and 
${\bf v}_{ij}={\bf v}_{i}-{\bf v}_{j}$.  
The rate of change of the elastic tangential displacement 
${\bf u}_{t_{ij}}$, set to zero at the initiation of a contact, 
is given by
\begin{equation}
\frac{d{\bf u}_{t_{ij}}}{dt}={\bf v}_{t_{ij}}
-\frac{({\bf u}_{t_{ij}}.{\bf v}_{ij}){\bf r}_{ij}}{r_{ij}^{2}}.
\label{equation4}
\end{equation}
The second term in
Eq.(\ref{equation4}) arises from the rigid body rotation around the
contact point and insures that ${\bf u}_{t_{ij}}$ always lies in the
local tangent plane of contact.  Normal and tangential forces acting
on particle $i$ are given by
\begin{eqnarray}
{\bf F}_{n_{ij}}&=&f(\delta_{ij}/d)(k_{n}\delta_{ij}{\bf n}_{ij}
-\gamma_{n}m_{\rm eff}{\bf v}_{n_{ij}}), \\
\label{equation5}
{\bf F}_{t_{ij}}&=&f(\delta/d)(-k_{t}{\bf u}_{t_{ij}}
-\gamma_{t}m_{\rm eff}{\bf v}_{t_{ij}}), 
\label{equation6}
\end{eqnarray}
where $k_{n,t}$ and $\gamma_{n,t}$ are elastic and viscoelastic
constants respectively, and $m_{\rm eff}=m_{i}m_{j}/(m_{i}+m_{j})$ is
the effective mass of spheres with masses $m_{i}$ and $m_{j}$. 
The corresponding contact
force on particle $j$ is simply given by Newton's third law, i.e., 
${\bf F}_{ji}=-{\bf F}_{ij}$.
For spheres of equal mass $m$, as is the case here, $m_{\rm eff}=m/2$;
$f(x)=1$ for the linear spring-dashpot model, denoted henceforth 
as Model L, or $f(x)=\sqrt{x}$ for Hertzian contacts with viscoelastic
damping between spheres, denoted as Model H. 

Our results are given in non-dimensional quantities by defining the
following normalization parameters: distances, times, velocities,
forces, elastic constants, and stresses are respectively measured in
units of
$d,~t_{o}=\sqrt{d/g},~v_{o}=\sqrt{gd},~F_{o}=mg,~k_{o}=mg/d,~\text{
  and }~\sigma_{o}=mg/d^{2}$. For a realistic simulation of glass
spheres with diameter $d=100\mu m$, the appropriate elastic constant
$k^{glass}_{n}=O(10^{10})$ necessitates a very small time-step
for accurate simulation, prohibiting any large-scale study. In
our simulations, we typically use a value for $k_{n}=O(10^{5})$ which
we believe captures the general behavior of intermediate-to-high-$k$
systems, thus offering a reasonable representation of realistic
granular materials (we discuss this aspect further in Section
\ref{intparam}).
A complete list of model parameters used in our standard simulation 
set, which consists
of 2D and 3D versions of Model L (L2 and L3), and a 3D version of 
Model H (H3), are given in Table~\ref{table1}. 

In a gravitational field {\bf g}, the translational and rotational
accelerations of particles are determined by Newton's second law,
in terms of the total forces and torques on each particle $i$:
\begin{eqnarray}
{\bf F}_{i}^{\rm tot}&=&m_i {\bf g} + \sum_{j}{{\bf F}_{n_{ij}}
+{\bf F}_{t_{ij}}}; \\
\btau_{i}^{\rm tot}&=&-\frac{1}{2}\sum_{j}{{\bf r}_{ij}
\wedge{\bf F}_{t_{ij}}}. 
\end{eqnarray}

The amount of energy lost in collisions is characterized by the
inelasticity through the value of the coefficient of restitution. For
Model L, there are separate coefficients, $\epsilon_{n}$ and
$\epsilon_{t}$, for the normal and tangential directions which are
related to their respective damping coefficients $\gamma_{n,t}$ and
spring constants $k_{n,t}$:
\begin{equation}
\epsilon_{n,t}=\exp(-\gamma_{n,t}t_{col}),
\end{equation}
where the collision time $t_{col}$ is given by,
\begin{equation}
t_{col}=\pi\left (2k_{n}/m - \gamma_{n}^{2}/4\right )^{-1/2}.
\end{equation}
The value of the spring constant should be large enough to avoid grain
interpenetration, yet not so large as to require an unreasonably small
simulation time step $\delta t$, since an accurate simulation typically
requires $\delta t\sim t_{col}/50$. For Model H, the effective 
coefficients of restitution depend on the initial velocity of the 
particles.

The static yield criterion, characterized by a local
particle friction coefficient $\mu$ \cite{footnote2}, is modeled by
truncating the  magnitude of ${\bf   u}_{t_{ij}}$ as necessary to satisfy 
$|{\bf F}_{t_{ij}}|<|\mu{\bf F}_{n_{ij}}|$. 
Thus the contact surfaces are treated as ``stuck'' while $F_{t_{ij}}<\mu
F_{n_{ij}}$, and as ``slipping'' while the yield criterion is satisfied. 
This ``proportional loading'' approximation \cite{deniz1} is a 
simplification of the much more complicated and hysteretic behavior 
of real contacts \cite{mindlin1}. 
To test the robustness of the proportional loading assumption,
we also carried out
simulations with Model L in  which ${\bf u}_{t}$ is not truncated but the
local yield criterion $F_{t}<\mu F_{n}$, is implemented. 
Note that we do not believe this to be a physically reasonable 
choice. Results for 
the two cases are similar, although the average kinetic energy is 
somewhat smaller (by approximately $18\%$ for Model L2) 
when $u_{t}$ is
truncated compared to those simulations when $u_{t}$ is unbounded.

The components of the stress tensor $\sigma_{\alpha\beta}$
within a given sampling volume $V$ are computed as the sum over all 
particles $i$ within that sampling volume of the contact 
stress (virial) and kinetic terms,
\begin{equation}
\sigma_{\alpha\beta}=\frac{1}{V}\sum_{i} \left[
\sum_{j\neq i}\frac{{r}_{ij}^{\alpha} F_{ij}^{\beta}}{2}+m_i(v_i^{\alpha}-
\overline{v^{\alpha}})(v_i^{\beta}-\overline{v^{\beta}})\right],
\label{stresseq1}
\end{equation}
where $F_{ij}^{\beta}=F_{n_{ij}}^{\beta}+F_{t_{ij}}^{\beta}$, 
and $\overline{\bf v}$ is the time-averaged velocity of
the particles within the sampling volume $V$. The time-averaged 
velocity must be subtracted since the kinetic portion of the 
stress tensor is entirely due to fluctuations in the velocity field.

For Hertzian contacts \cite{johnson1}, the ratio $k_{t}/k_{n}$ depends
on the Poisson ratio of the material, and is about 2/3 for most
materials.  For ease in our simulations, we use a value
$k_{t}/k_{n}$=2/7, which makes the period of normal and shear
contact oscillations equal to each other for Model L \cite{wolf1}. 
However, the contact dynamics are not very sensitive to the 
precise value of this ratio. We have performed simulations with 
different
values of $k_{t}/k_{n}$ to test how this ratio may affect our results;
different values of this ratio yield nearly identical results. The
only difference we observe is a slight increase in the total, averaged kinetic
energy (KE) of the system when $k_{t}/k_{n}>2/7$, and a decrease
for $k_{t}/k_{n}<2/7$. For example,
when we set $k_{t}/k_{n}=2/3$ instead of $2/7$, the total averaged KE
increases by about $10\%$, whereas all other macroscopic quantities
measured in the simulations, such as density and stress, remain 
essentially unchanged. 

Similarly, although all results reported here are
for $\gamma_{t}/\gamma_{n}=0$ ({\it i.e.} no rotational velocity
damping term) we have also carried out simulations to measure the
effect of introducing rotational damping, $\gamma_{t}/\gamma_{n}>0$.
When we set $\gamma_{t}=\gamma_{n}$, we observe a slight decrease, of
about $8\%$, in the total averaged KE, compared with those
simulations that have $\gamma_{t}=0$. 
Making $\gamma_{t}/\gamma_{n}$
non-zero quickens the approach to the steady state by
draining out more energy. However, all other quantities are, again,
unchanged. We discuss reasons why we observe minimal changes with
these interaction parameters in Sec.~\ref{intparam}.

Typical values for the friction coefficient $\mu$ range between, 0.4
and 0.6 for many materials. We chose $\mu=0.50$ for
most of our simulations, though variations in $\mu$ will be 
discussed in Sec.~\ref{intparam}.
Similarly, the value of $\epsilon$ is chosen to reflect the properties
of a realistic granular particle.
\begin{table} 
\begin{center}
\begin{tabular}{ccccccccc}
Model & $D$ & $f(x)$ & $k_{n}$ & $\gamma_{n}$ & $k_{t}/k_{n}$ &
$\gamma_{t}/\gamma_{n}$ & $\mu$ & $\epsilon$ \\ \hline 
L2 & 2 & 1 & $2\cdot10^{5}$ & $33.5$& 2/7 & 0 & 0.50 & 0.92 \\ 
L3 & 3 & 1 & $2\cdot10^{5}$ & $50.0$& 2/7 & 0 & 0.50 & 0.88 \\ 
H3 & 3 & $\sqrt{x}$  & $2\cdot10^{5}$ & $50.0$& 2/7 & 0 & 0.50 & - 
\end{tabular} 
\caption{\em Parameter values used in our standard simulation set 
for the 2D and 3D linear spring models (Models L2 and L3, $f(x)=1$), 
and the 3D Hertzian model (Model H3, $f(x)=\sqrt{x}$). 
For Model H3, $\epsilon$ is velocity dependent.}
\label{table1}
\end{center}
\end{table}

The equations of motion for the translational and rotational degrees
of freedom are integrated with either a third-order Gear
predictor-corrector or velocity-Verlet scheme \cite{allen1} with a
time-step $\delta t=10^{-4}$ for $k_{n}=2\times 10^{5}$. All data was
taken after the system had reached the steady state. To reach the
steady state, simulations were required to run for $~ 1-2\times
10^{7}\delta t$ when starting from a non-flowing state for $N <
10,000$, and the largest system in 2D $(H=200)$ required a run time of
$~ 2-5\times 10^{8}\delta t$. On a 500 MHz DEC Alpha processor, our code
requires about 5 days to simulate 10 million timesteps of a 3D
8000-particle granular system. We have also created a parallel version
of the 3D code using the standardized MPI message-passing library.
The parallel code partitions the simulation domain into small 3D
sub-blocks using the methods described in \cite{plimpton1}. Even on a
cluster computer with relatively low interprocessor communication
bandwidth, the code runs at high parallel efficiencies so long as we
simulate 1000 or so particles per processor. For example, on 8
processors of our Alpha/Myrinet cluster, we can simulate 15 million
timesteps/day of the same 8000-particle system.

For the imposition of chute flows with varying tilt angles, we rotate
the gravity vector ${\bf g}$ in the $xz$-plane by the tilt angle
$\theta$ away from the $-{\bf z}$ direction --- flow is from left to
right in this sense. This means that the ${\bf z}$-axis is always
normal to the free surface.  In 3D the area of the bottom is
$A=L_{x}L_{y}$ where $L_{x}$ and $L_{y}$ are the dimensions of the
simulation cell in the $x$ and $y$ directions respectively. For the 3D
simulations, we define a measure of the height of the pile by defining
$H\equiv Nd^{2}/A$ as the pile height if it were sitting on a level
plane at rest in a simple cubic lattice.  For example, for $N=8000$
and $L_{x}=20d$ and $L_{y}=10d$, $H=40$ (although due to the precise
configuration, the actual measured height $\approx 37$). This is a
useful definition for comparing between different system sizes.  We
study a range of system sizes, $1000\le N\le 20000$.  For the largest
system, $H=100$. The influence of other wall dimensions $L_{x},~L_{y}$
was also studied. For the 2D runs, the $x$-dimension of the periodic
side is fixed at $100d$ ({\it i.e.} $50$ large particles long) and the
pile height $2\leq H \leq200$, {\it i.e.}  $N=200-20000$.

\begin{figure}
\begin{center}
\begin{tabular}{cc}
  \resizebox{!}{9cm}{\includegraphics*[bb=0.0cm 5cm 19cm
    24cm]{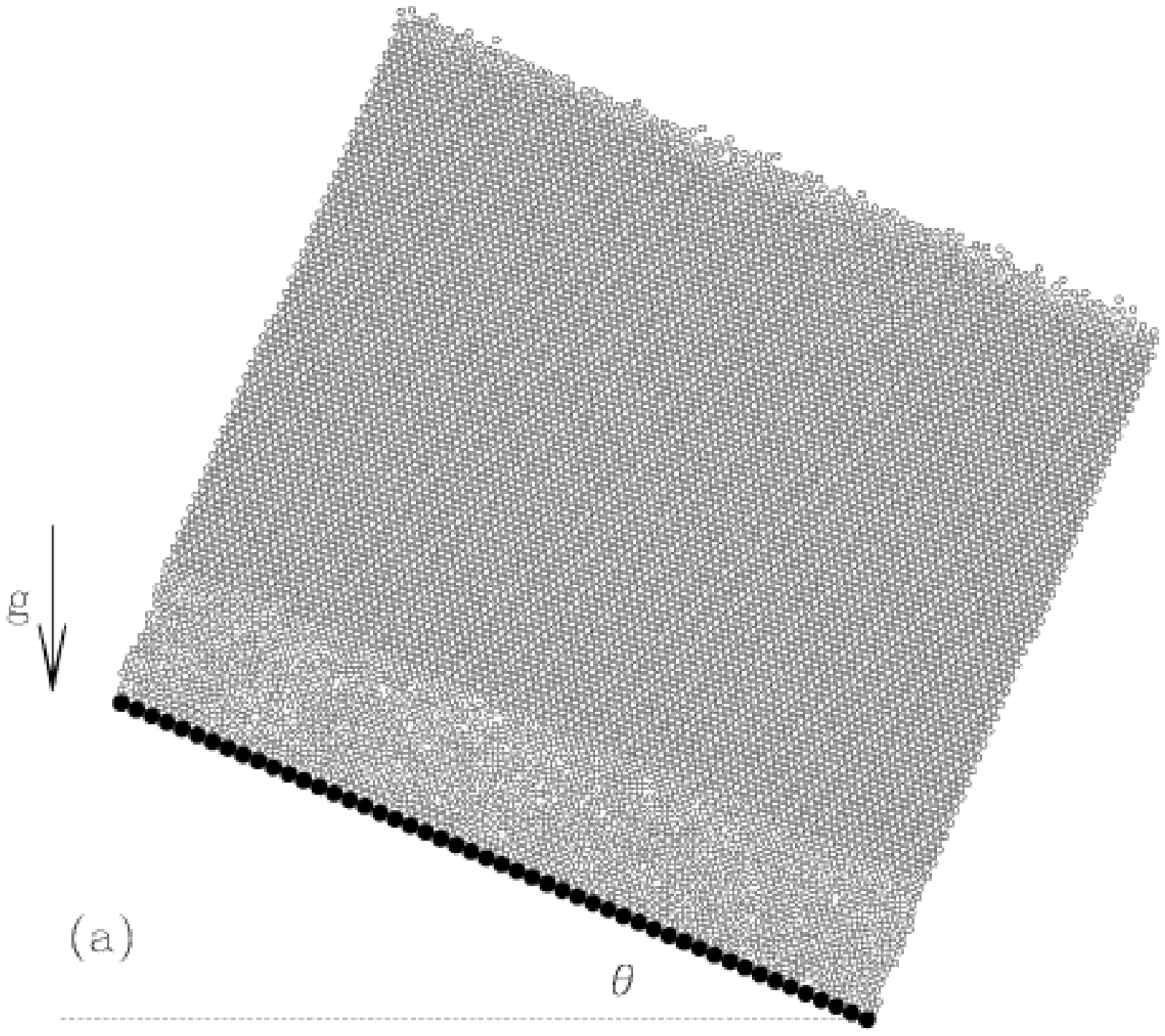}}
  \resizebox{!}{9cm}{\includegraphics*[bb=0.0cm 5cm 19cm
    24cm]{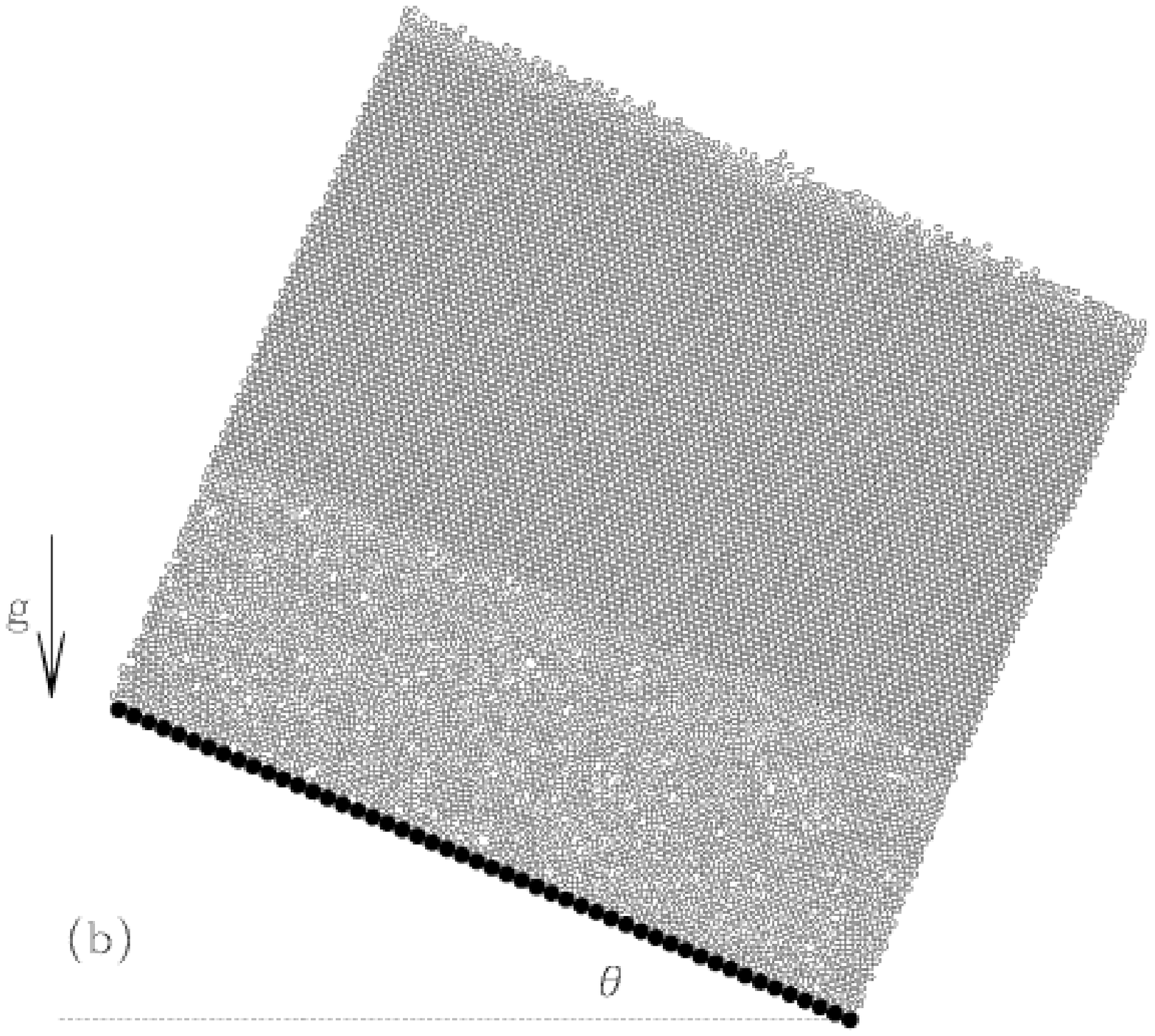}}\\
  \resizebox{!}{9cm}{\includegraphics*[bb=0.0cm 5cm 19cm
    24cm]{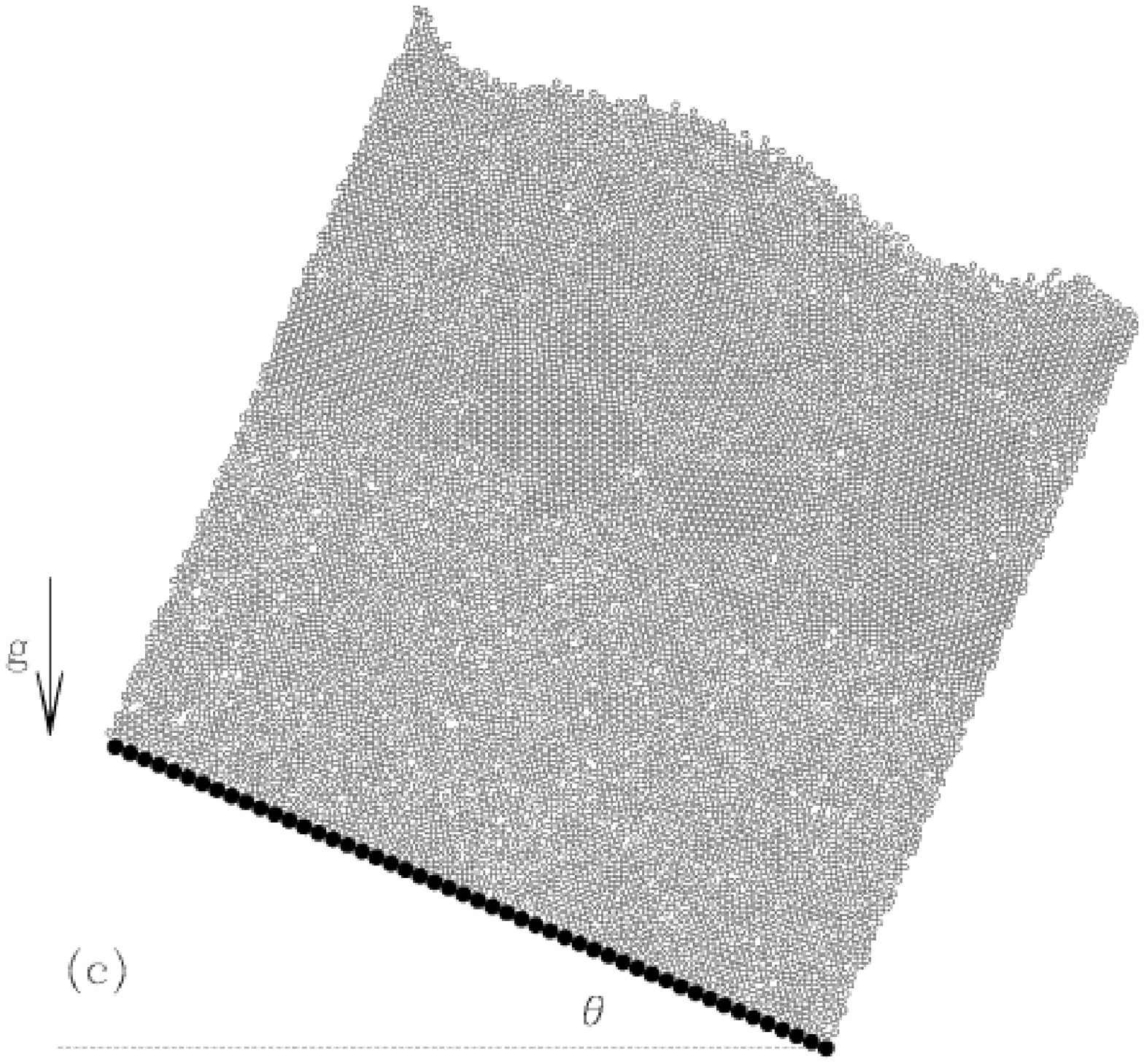}}
  \resizebox{!}{9cm}{\includegraphics*[bb=0.0cm 5cm 19cm
    24cm]{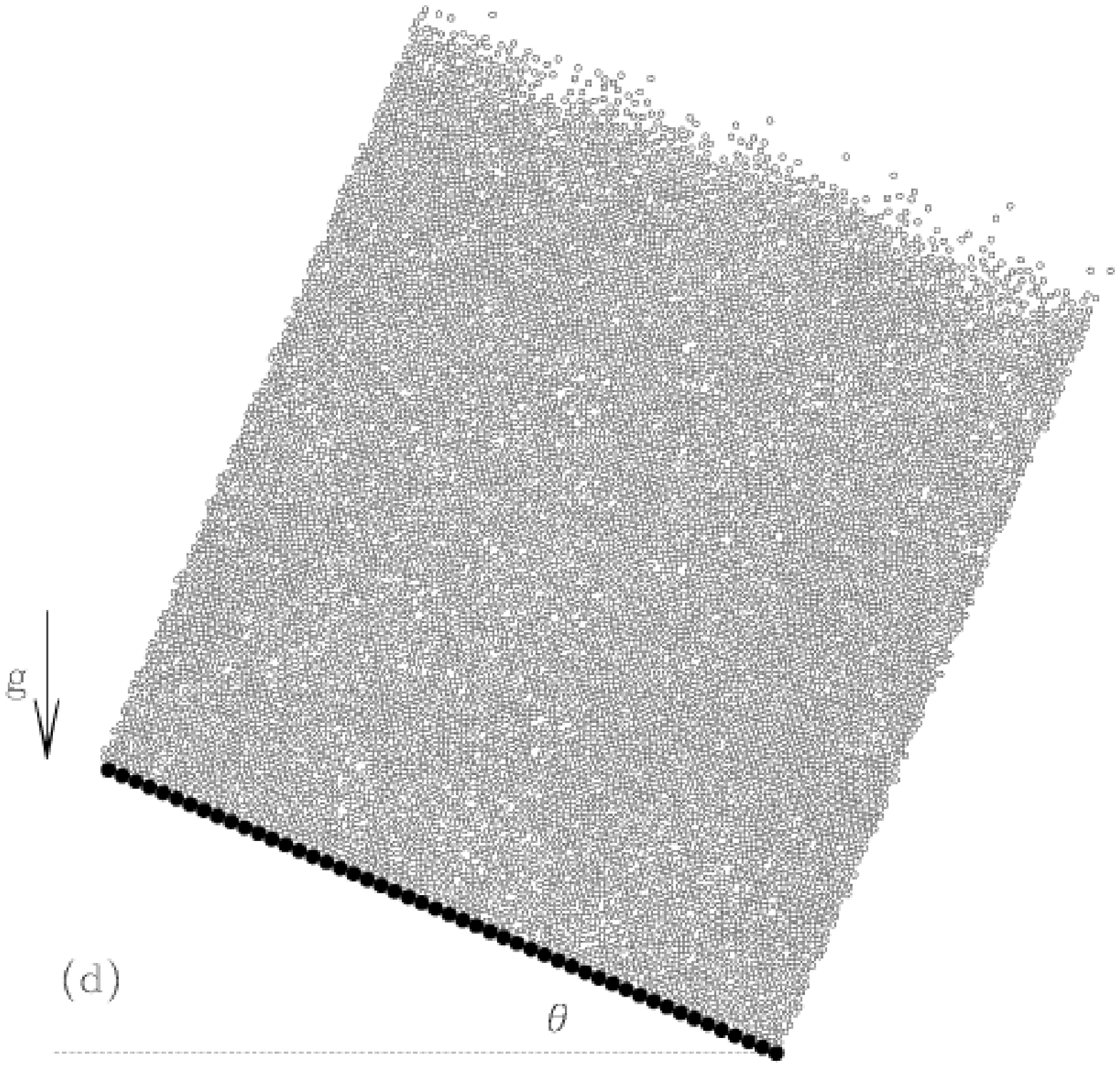}}
\end{tabular}
\caption{\em Time sequence of a typical configuration in 2D following an 
instantaneous change in the inclination angle $\theta$ from $0^{o}$ to
$24^{o}$. Results are for $N=10000,~\mu=0.50$, and $\epsilon=0.82$,
and for times $t$ = a) 100, b) 400, c) 600, d) 6000. 
Flow is left to right. As the flow progresses, the dilational front 
propagates upwards through the system, destroying the initial ordered 
array; the pile consequently ``fluffs'' up.} 
\label{config_2d} 
\end{center} 
\end{figure}

In 2D, the initial state was constructed by building a triangular
lattice of particles.  The tilt angle was then increased until flow
occurred. The initial flow occurred only for $\theta\gtrsim 23^{o}$.
This minimum value to induce flow depends on the size and spacing of
bottom plate particles. The initial failure occurred mostly at the
bottom of the pile, followed by movement of a dilation front toward
the top of the pile as shown in Fig.~\ref{config_2d}. Once this
initial steady state was achieved, the angle $\theta$ was adjusted to
its desired value, and the system equilibrated to its final steady
state. In 3D, we started the system from a randomly diluted simple
cubic lattice. The angle was then increased to a large angle
$\theta=30^{o}$ to induce disorder and settling of particles. The
angle $\theta$ was then decreased to the desired value and flow
allowed to continue until a steady state was reached, before
measurements were taken.

In 3D, to test for hysteresis near $\theta_{r}$, $\theta$ was reduced
to below $\theta_{r}$ until the system settled down into a disordered
state and stopped flowing. $\theta$ was subsequently increased to
$\theta_{r}^{flow}$ and the system began flowing. This angle of flow
initiation was sometimes different from the angle of cessation of flow
$\theta_{r}^{stop}$ when taking a flowing state and then lowering
$\theta$ down to $\theta_{r}^{stop}$ to stop the flow. However, this
small hysteretic behavior, in 3D, only occurs for thin piles at low
angles. 

In 2D the equivalent phase diagram can only be constructed by
taking a flowing state at angle $\theta$ and then lowering to
$\theta_{r}^{stop}$. Once the 2D state stops flowing the system
spontaneously crystallizes into a polycrystalline ordered state. To
induce flow from this ordered state requires increasing $\theta$ to a
much higher angle than $\theta_{r}^{stop}$.

\section{Results: Velocity and Density Profiles}
\label{results}

\subsection{Kinematics of steady state systems}
\label{steady}

\begin{figure}
\begin{center}
\includegraphics[width=2.9in]{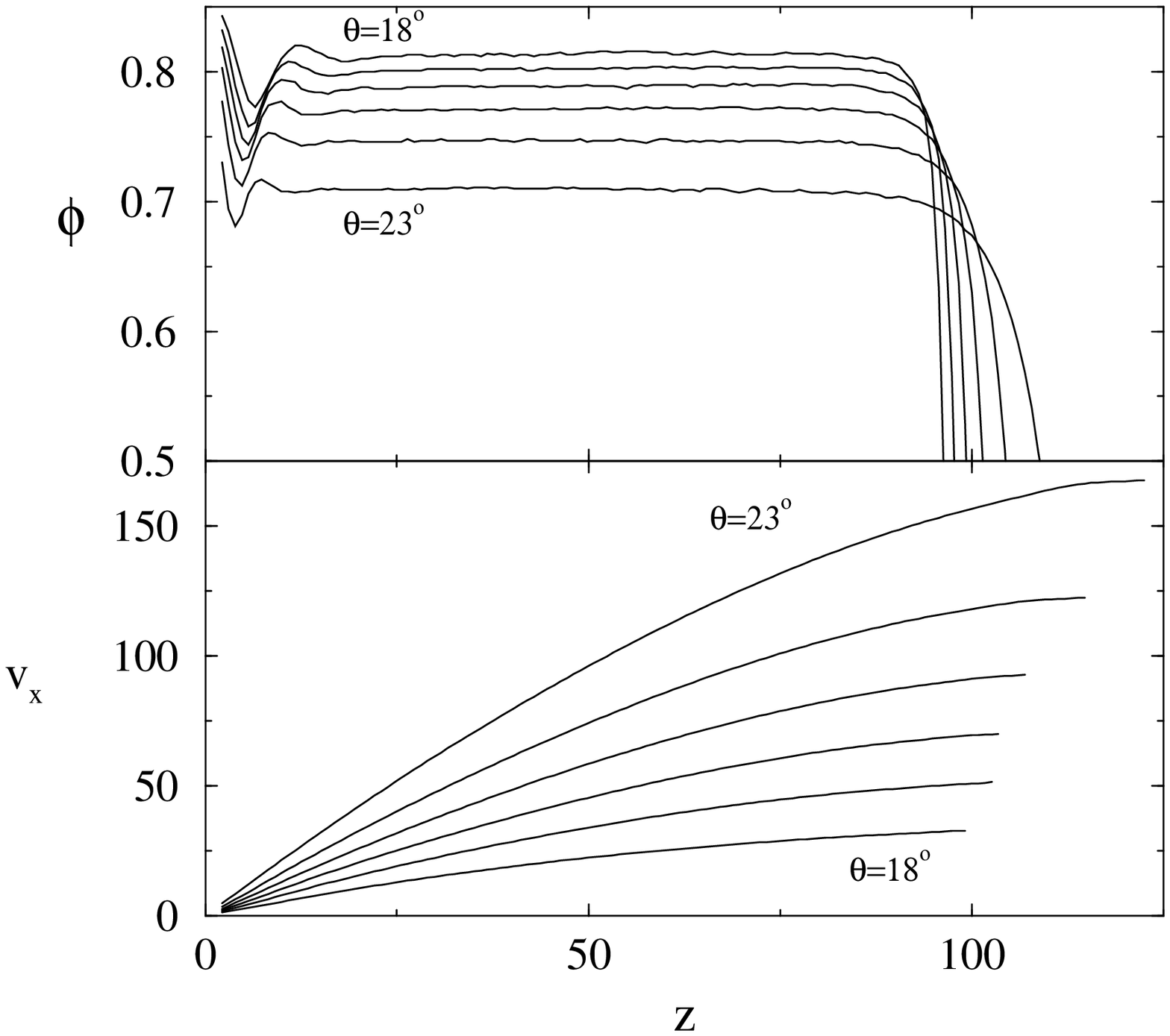}\hfil
\includegraphics[width=2.9in]{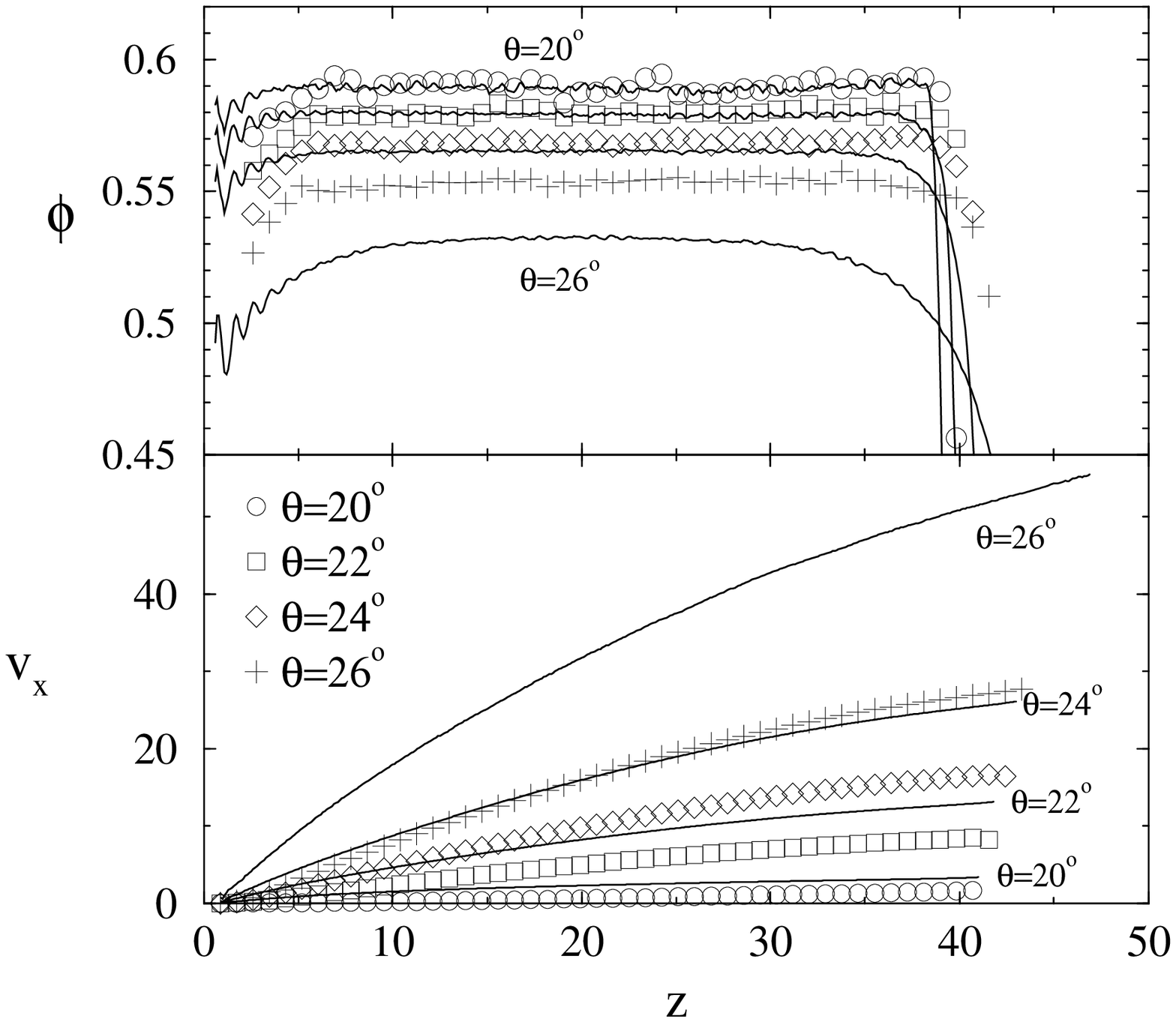}\\
\hfil(a)\hfil\hfil(b)\hfil\\
\bigskip
\caption{\em Packing fraction $\phi$ and velocity $v_x$ profiles,
as a function of distance from bottom $z$, for (a) 2D spring--dash-pot 
model (Model L2), with $H=100$, at tilt angles of 
$\theta$=18, 19, 20, 21, 22, 23 degrees. (b) 3D, $H=40$ systems at
$\theta$=20, 22, 24, 26 degrees, with Model L3 (open symbols) and 
Model H3 (solid lines).} 
\label{denandvel1} 
\end{center}
\end{figure}

We focus our main
attention on the regime of steady state flow for moderate to deep
piles, for which $\theta_{r}$ is independent of depth. In
Fig.~\ref{denandvel1} we plot the density and velocity (in the
direction of flow) $z$-profiles over a range of inclination angles
$\theta$, for a series of simulations in 2D and 3D. 
Figure~\ref{denandvel1}(a) is for a 2D system (Model L2, cf. 
Table~\ref{table1}) of N=10,000 particles, corresponding to
$H=100$. In  Fig.~\ref{denandvel1}(b), the equivalent 3D
model (Model L3 with $N=8,000, H=40$), 
denoted by the open symbols, is compared to the 3D Hertzian 
model (Model H3). The tilt angle was varied between
$18^{o}-30^{o}$ in all cases. In 2D the system becomes unstable 
to an accelerating flow above $23^{o}$ and in 3D the unstable flow 
regime is observed above $26^{o}$. 

In both 2D and 3D, the packing fraction 
remains constant over almost 40 layers in 3D, and 100 layers in 2D. 
For all steady state systems, as the 
tilt angle is increased, the value of the
bulk packing fraction decreases. This decrease
accompanies a growing dilated region (of lower packing fraction)
near the free surface at the top. All the velocity
profiles are concave, and velocities increase in value with increasing 
tilt angles.
Consequently, the total kinetic energy of the system rises with
increasing angle \cite{footnote1}.

\end{multicols}

\begin{multicols}{2}

\begin{figure}
\begin{center}
\includegraphics[width=2.9in]{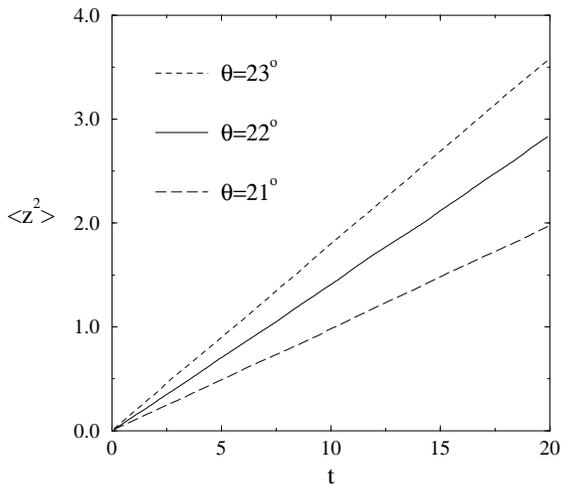}
\caption{\em z-component of the mean square displacement for 3 angles, 
$\theta=21^{o}, 22^{o}$, and $23^{o}$, for Model L3, with
$H=40$.} 
\label{diffusion1}
\end{center}
\end{figure}

We monitor vertical mixing of the bulk by measuring
the bulk-averaged, mean-square displacement of particles over time.
Figure~\ref{diffusion1} shows the mean-square displacement 
of particles normal to the surface $<z^{2}>$ as a function of
simulation time for Model L3, over a range of tilt angles. The linear
relationship demonstrates well-defined diffusive motion in the
z-direction, suggesting thorough mixing in the system.
Similar results are observed in 2D.  At long times $<z^{2}>$ will
reach a constant due to the finite height of the pile.

\begin{figure}
\includegraphics[width=2.9in]{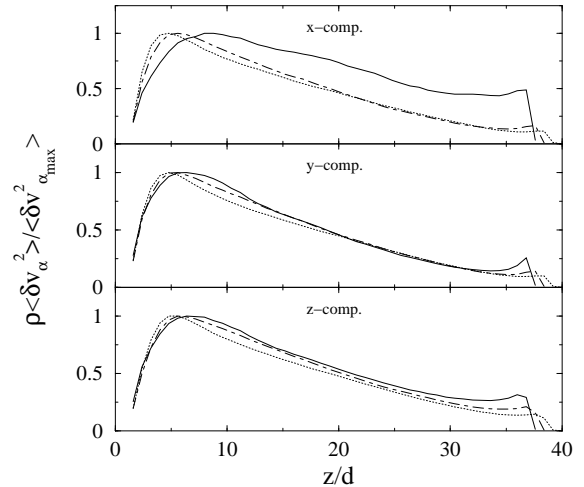}
\caption{\em Profiles of the kinetic portion of the diagonal
elements of the stress tensor 
$\rho<(\delta v^{\alpha})^{2}>$, normalized by their maximum value
along the curve,  for Model L3 inclined at 
$21^{o}~(- - -)$, $23^{o}~($---$)$, and
$25^{o}~(- \cdot)$.} 
\label{velflucs}
\end{figure}

By observing a sequence of snapshots (not shown here) of tracer
particles at various heights in the bulk, we also find that diffusion
is somewhat faster nearer the bottom of the pile. 
This is indicative of the fact that fluctuations in the particle
velocities are greater closer to the bottom wall. 
Fig.~\ref{velflucs} depicts the diagonal components of the kinetic part
of the stress tensor, $\rho<(\delta v^{\alpha})^2>$, 
where $\rho$ is the mass density and 
$\delta v^{\alpha}=v^{\alpha}-\overline{v^{\alpha}}$, at three different
angles for Model L3. Indeed we do find that the velocity
fluctuations (frequently termed ``granular temperature'' in the 
literature of dilute granular flows) are greatest at the bottom 
of the pile (away from the actual plate) and decrease with height 
until the values appear to level off at the top free surface.

This behavior partially illustrates how the pile is able to maintain
a constant density profile, even though the stresses increase towards 
the bottom, and the flowing pile has a finite compressibility, as 
evidenced by the changing density as a function of tilt angle $\theta$.
Particles deeper into the pile experience increasing compaction forces
due to the load of the particles above, yet a constant density 
is maintained through the increased particle velocity fluctuations.

\begin{figure}
\begin{center}
\includegraphics[width=2.9in]{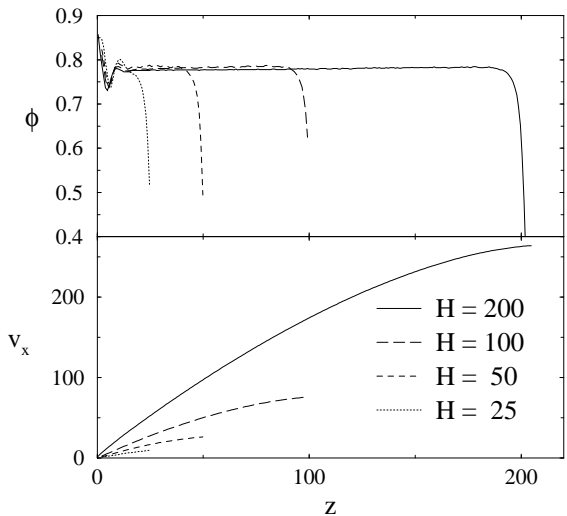}\hfil
\includegraphics[width=2.9in]{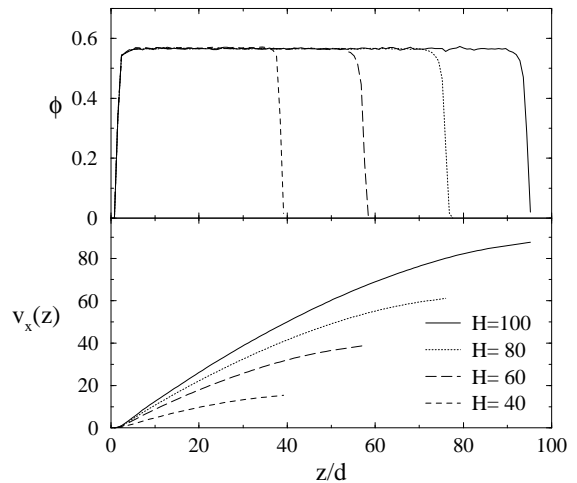}\\
\hfil(a)\hfil\hfil(b)\hfil\\
\bigskip
\caption{\em Density and velocity profiles for (a) 2D systems (Model L2)
for $\theta=20^{o}$ with sizes $H=200,~100,~50,~25$ and (b) 3D systems (Model L3) for $\theta=24^{o}$ with sizes $H=100,~60,~50,~60$ .} 
\label{denandvel2}
\end{center}
\end{figure}

The data sets shown in Fig.~\ref{denandvel1} are for one system size
only. In Fig.~\ref{denandvel2} density and velocity profiles for
systems of varying heights are compared. The densities measured deep in
the pile, as well as the density and strain rate profiles near the 
surface, are independent of the overall height of the pile.
This suggests that the rheology 
of the system is local in this regime; i. e., that constitutive relations
locally relate stress and strain rate. For reasons alluded to in 
Sec.~\ref{secrheo}, we have been unable to identify these constitutive
relations. 

\end{multicols}

\begin{multicols}{2}
\begin{figure}
\begin{center}
\includegraphics[width=2.9in]{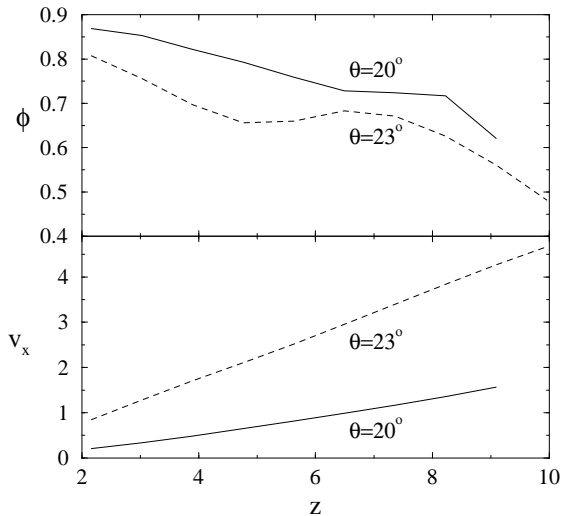}
\caption{\em Density and velocity profiles for thin systems (Model L2),
with $H=10$, at $20^{o}$ and $23^{o}$. 
These profiles are very different from the thicker piles.} 
\label{denandvelsmall} 
\end{center}
\end{figure}

We note that the behavior observed in Figs.~\ref{denandvel1} and
\ref{denandvel2} is true only for $H\gtrsim 20$. For smaller
piles, such as those commonly studied experimentally \cite{hungr1},
the behavior is very different, as seen in Fig.~\ref{denandvelsmall},
where a linear velocity profile is evident. Recent experimental
studies of thin piles in inclined plane geometries also report linear
velocity profiles \cite{durian1}.

\subsection{Dependence on Interaction Parameters}
\label{intparam}
In this subsection, we investigate the sensitivity of these
results to the particle interaction parameters. We independently
vary the internal coefficient of friction $\mu$, the coefficient of
restitution $\epsilon$, and the value of the spring constant
$k_{n}$. We observe that while the density of the
bulk material does not depend sensitively on these interaction
parameters, the velocity profiles do. 

\begin{figure}
\begin{center}
\includegraphics[width=2.9in]{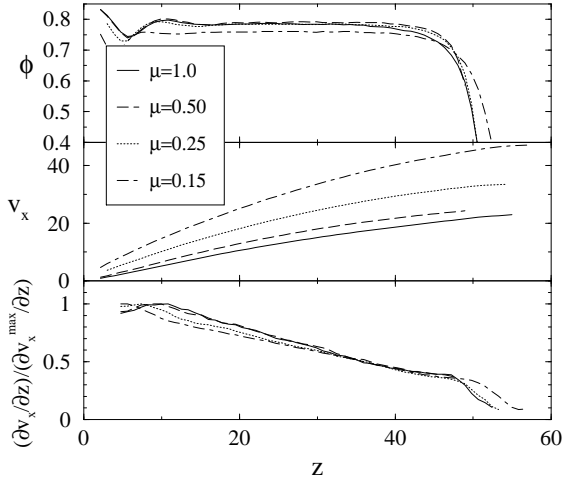}\hfil
\includegraphics[width=2.9in]{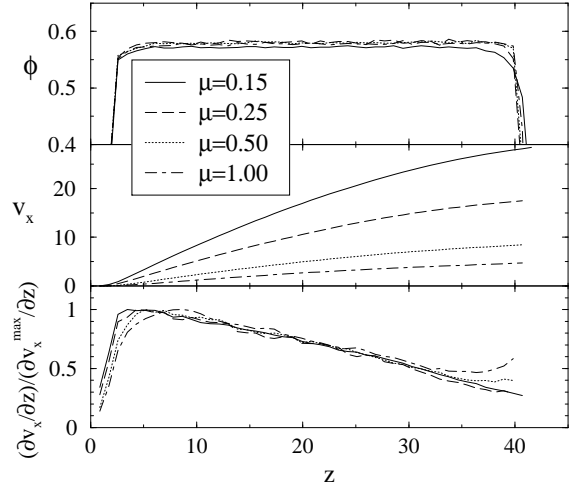}\\
\hfil(a)\hfil\hfil(b)\hfil\\
\bigskip
\caption{\em Density, velocity, and strain rate profiles for different 
values of the particle friction coefficient for:
(a) Model L2 at $\theta=20^{o}, H=50$ for $\mu=0.15,~0.25,~0.50,~1.0$, 
and (b) Model L3 at $\theta=22^{o}, H=40$ for
$\mu=0.15,~0.25,~0.50,~1.0$.} 
\label{denandvel3} 
\end{center}
\end{figure}

Figure~\ref{denandvel3} shows the sensitivity to the friction 
coefficient $\mu$ by depicting density, velocity, and strain rate
profiles for: (a) Model L2 with $H=50$ and $\theta=20^{o}$,
where $\mu=0.15,~0.25,~0.50,~1.0$, and 
(b) Model L3 with $H=40$ and $\theta=22^{o}$, for values of
$\mu=0.15,~0.25,~0.5,~1.0$. The data suggest that there is minimal
change in the bulk density over this range in $\mu$. In the bottom
panels of Fig.~\ref{denandvel3}, the shear rate $\frac{\partial
  v_{x}}{\partial z}$ scaled by $\frac{\partial v_{x}^{max}}{\partial z}$ is
plotted for the various values of $\mu$.

\end{multicols}

\begin{figure}
\begin{center}
\includegraphics[width=2.9in]{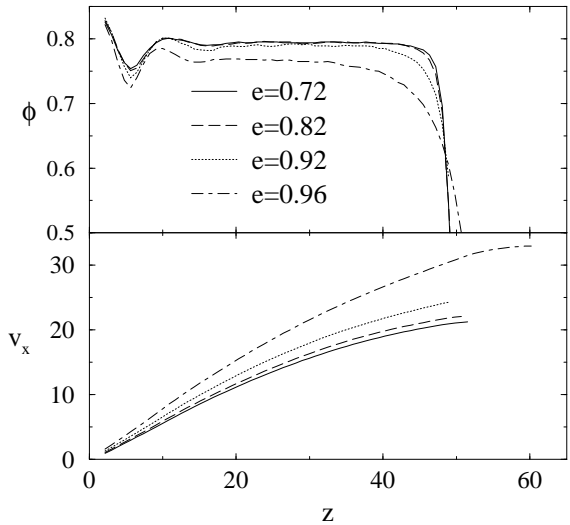}\hfil
\includegraphics[width=2.9in]{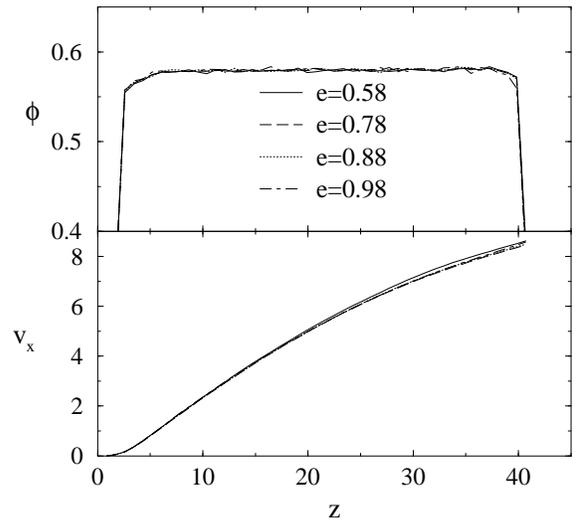}\\
\hfil(a)\hfil\hfil(b)\hfil\\
\bigskip
\caption{\em Density and velocity profiles for different values of 
$\epsilon$ for 
(a) Model L2 at $\theta=20^{o}, H=50$ for $\epsilon=0.72, 0.82, 0.92,
0.96$, and (b) Model L3 at $\theta=22^{o}, H=40$ for 
$\epsilon=0.58, 0.78, 0.88, 0.98$.} 
\label{denandvel4} 
\end{center}
\end{figure}

\begin{multicols}{2}
Similarly, Fig.~\ref{denandvel4} shows the profiles for the 
same systems as described in Fig.~\ref{denandvel3}, 
but with a fixed $\mu=0.5$ and varying coefficients of restitution
$\epsilon$, {\it i.e.}, varying the inelasticity of the system. 
Again we see that variations in
$\epsilon$ have little effect on the flow behavior of these systems,
particularly in 3D, provided that 
the system is able to reach steady state. [For low 
$\mu$ ($\approx 0.10$) and high $\epsilon$ 
($\approx$ 0.96 for 2d and 0.98 for 3d), the systems become
unstable.]
\end{multicols}

\begin{figure}
\begin{center}
\includegraphics[width=2.9in]{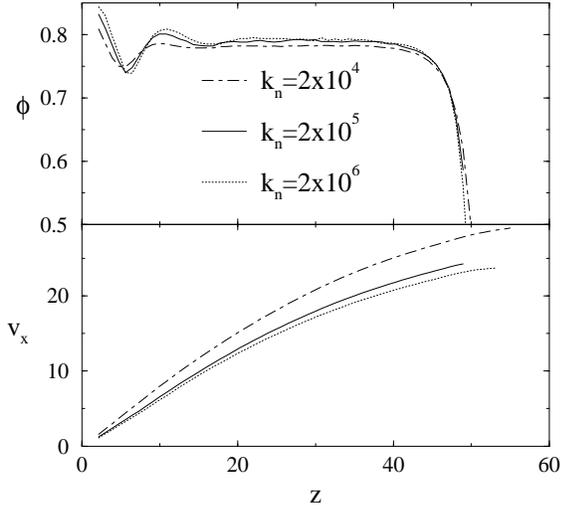}
\caption{\em Density and velocity profiles for different values of 
the spring constant $k_{n}$, for Model L2 at $\theta=20^{o}, H=50$.} 
\label{denandvel5}
\end{center}
\end{figure}

\begin{multicols}{2}
Another microscopic parameter we have investigated is the effective 
hardness of the particle, determined by the value of the spring 
constant $k_{n}$. We vary $k_{n}$
and keep $\epsilon$ constant by adjusting the value of $\gamma_{n}$.
Simulations investigating this parameter can be time-consuming:
increasing $k_{n}$ by a factor of 100 requires a reduction in the
time-step by a factor of 10.  Fortunately, as Fig.~\ref{denandvel5},
(measured for Model L2 with $\theta=20^{o}, H=50$) indicates, 
the effect of variations in $k_{n}$ is minimal, provided 
$k_{n}$ is sufficiently large.

\begin{figure}
\begin{center}
\includegraphics[width=2.9in]{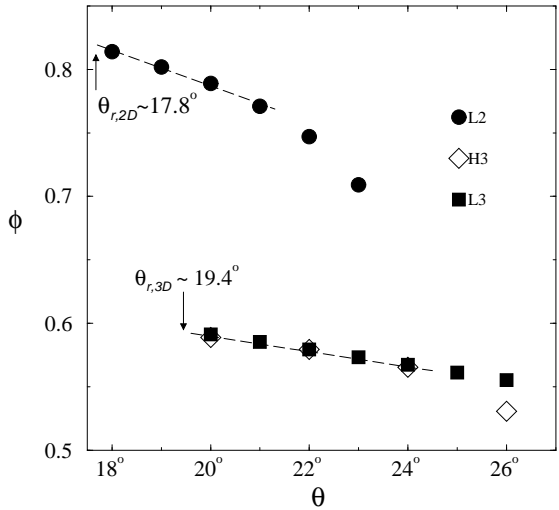}
\caption{\em Tilt dependence of the packing fraction in the region of 
constant packing fraction, for Models L2 (solid circles), L3 (solid squares),
and H3 (open diamonds). The dashed lines denote the linear dependence
on tilt angle near the angle of repose.}
\label{denvsangle} 
\end{center}
\end{figure}

\subsection{Dependence on Tilt Angle}
Judging by the insensitivity of the macroscopic quantities to the various
interaction parameters for Model L (as shown), as well as Model H, 
we see that to a good approximation, effects due to material properties 
and system size can be neglected in the steady state regime. 
As shown in
Fig.~\ref{denvsangle}, the packing densities vary approximately 
linearly with $\theta$ and approach the maximum values 
$\phi_{2D}^{\rm max}=0.815(5)$ 
and $\phi_{3D}^{\rm max}=0.590(5)$ at $\theta_{r,2D}\approx 17.8^\circ $ 
and $\theta_{r,3D}\approx 19.4^\circ $ for 2D and 3D, respectively.

It is interesting to note that the asymptotic packing
fractions $\phi^{max}_{2D}$ and $\phi^{max}_{3D}$ are close to
the values one would obtain assuming the flow was the densest possible
flow of lines (in 2D) or planes (3D) of close-packed particles parallel 
to the top surface.  
For the 2D case, the packing corresponds to a square lattice with
a packing fraction of $\pi/4 \simeq .79$.  For the 3D case, the
sliding planes would be square lattices, stacked to form triangular 
lattices in the $y-z$ plane. This arrangement has a packing fraction 
of $\frac{\pi}{3 \sqrt{3}} \simeq .60$.  

\section{Results: Stress Analysis}
\label{secstress}
\subsection{Cauchy Equations}
The stress tensor is symmetric: $\sigma_{ij}=\sigma_{ji}$, 
with $D(D+1)/2$ independent
components in $D$-dimensional space. The Cauchy (force-balance) 
condition provides only $D$ equations, leaving the solution 
underdetermined.  Thus, an additional $D(D-1)/2$
 constitutive relations are needed to close the equations and to solve 
for the transmission of stress in a granular system.  

In 2D, the steady state Cauchy equations are
\begin{eqnarray}
\frac{\partial \sigma_{zz}}{\partial z} &=& \rho g \cos\theta, \\
\frac{\partial \sigma_{xz}}{\partial z} &=& \rho g \sin\theta.
\end{eqnarray}
For a given tilt angle, these give: 
\begin{equation}
\sigma_{zz}(z)=g\cos\theta \int_{z}^{\infty}dz\rho(z)
\label{zzequalweight}
\end{equation}
\begin{equation}
\sigma_{xz}(z)=\sigma_{zz}(z)\tan\theta,
\label{xzequalzz}
\end{equation}
where $\rho$ is the number density of spheres ($\phi=\pi\rho
d^{D}/2D$ for dimensionality $D=$ 2 and 3).  If, as in our case, the
density $\rho$ is constant,
\begin{eqnarray}
\label{eqhdef}
\sigma_{zz}(z)&=& g\rho \cos\theta (h-z), \\
\label{eqsxz}
\sigma_{xz}(z)&=& g\rho \sin\theta (h-z), 
\end{eqnarray}
\noindent 
where $h$ is the effective height of the flowing pile, which 
appears as a constant of integration in Eq.(\ref{eqhdef}).
$\sigma_{xx}$ cannot be
determined from these considerations; as we lack a constitutive 
relation that would determine it. 
Nevertheless, important features of the behavior of the stress tensor
can be obtained by Mohr-Coulomb analysis\cite{nedderman1}.

\subsection{Mohr-Coulomb Analysis}
The Mohr circle, shown in Fig.~\ref{figMohr1}a, is a 
geometrical construction that enables visualization of 
rotational transformations of the stress tensor. 
The circle is drawn in the $\sigma-\tau$ plane, 
such that the points $A~(\sigma_{zz},\sigma_{xz})$
and $B~(\sigma_{xx},-\sigma_{xz})$ form a
diameter of the circle, centered at point $O$.
Coordinates of the points on the circle represent the normal
($\sigma$) and shear ($\tau$) components of the stress tensor
associated with all possible shear planes. 
Upon a rotation of the coordinate system, i.e., the plane
in which shear is specified, by an angle $\psi$, the representative
points rotate by an angle $2\psi$ around the circle.  

At a given tilt angle, $\sigma_{zz}$ and $\sigma_{xz}$ are determined 
by the Cauchy equations, which fixes the location of point $A$
($\sigma_{xz}/\sigma_{zz}=\tan\theta$).  However, $\sigma_{xx}$, and
thus the location of point $B$, is undetermined by the Cauchy
equations, and depends on the rheology.

The ``stress angle" $2\varphi\equiv\widehat{COA}$, formed by the
stress point $A$, origin of the Mohr Circle $O$, and point $C$, whose
tangent passes through the origin of the ($\sigma,\tau$) plane, can be
used as a surrogate for any quantity that completes the description of
the $x-z$ stress state, since it uniquely identifies the 
two-dimensional stress state
of the flowing pile (for $\theta > \theta_r$) by fixing the value of
$\sigma_{xx}$.  For a pile with a uniform Coulomb yield criterion that
is at incipient yield everywhere (IYE) when $\theta=\theta_r$, the
points $C$ and $A$ coincide, and therefore $\varphi=0$. On the other
hand, if the flowing pile behaves like a fluid,
$\sigma_{xx}=\sigma_{zz}$, and consequently $\sin2\varphi=\tan\theta$.

\subsection{Stress Tensor Near the Surface}
\begin{figure}
\begin{center}
\includegraphics[width=2.9in]{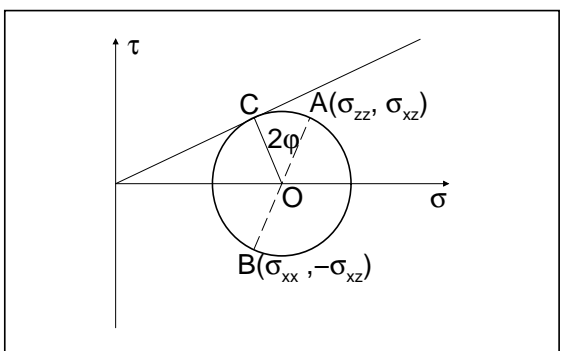}\hfil
\includegraphics[width=2.9in]{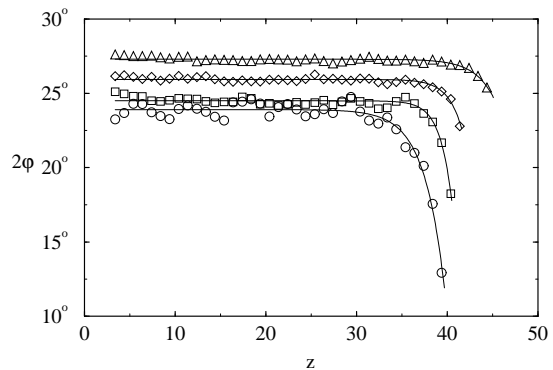}\\
\hfil(a)\hfil\hfil(b)\hfil\\
\bigskip
\caption{\em (a) The Mohr circle is a graphical tool that is
used to determine transformations of a rank 2 tensor (such as stress)
under rotation. The stress components for a given coordinate 
system are represented by points $A$ and $B$, which form a diameter
of the circle. The transformed stress components upon a rotation of 
the coordinate system by angle $\psi$ can be found by a rotation of
these points by $2\psi$ around the circle. The point $C$, which has
a tangent that passes through the origin, corresponds to the 
orientation of a shear plane (at an angle $\varphi$ to the $x-$axis)
with the largest ratio of shear to normal stress. 
  (b) The stress angle $2\varphi$
  [$\widehat{COA}$ in (a)] as a function of height for
  $\theta=20^\circ$ ($\bigcirc$), $22^\circ$
  ($\Box$), $24^\circ$ ($\Diamond$), and $26^\circ$ ($\triangle$).
  The results are for Model H3 with H=40. The lines are fits that
  decay exponentially from $2\varphi^{\rm surf}$ at the
  effective height $h$ [cf. Eq.(\protect\ref{eqhdef})] to 
  $2\varphi^{\rm bulk}$ in the bulk, with a 
  typical decay length of 1.3 to $2.2d$, indicating a
  surface layer about $5d$ to $8d$ thick.}
\label{figMohr1}
\end{center} 
\end{figure}

\begin{figure}
\begin{center}
\includegraphics[width=2.9in]{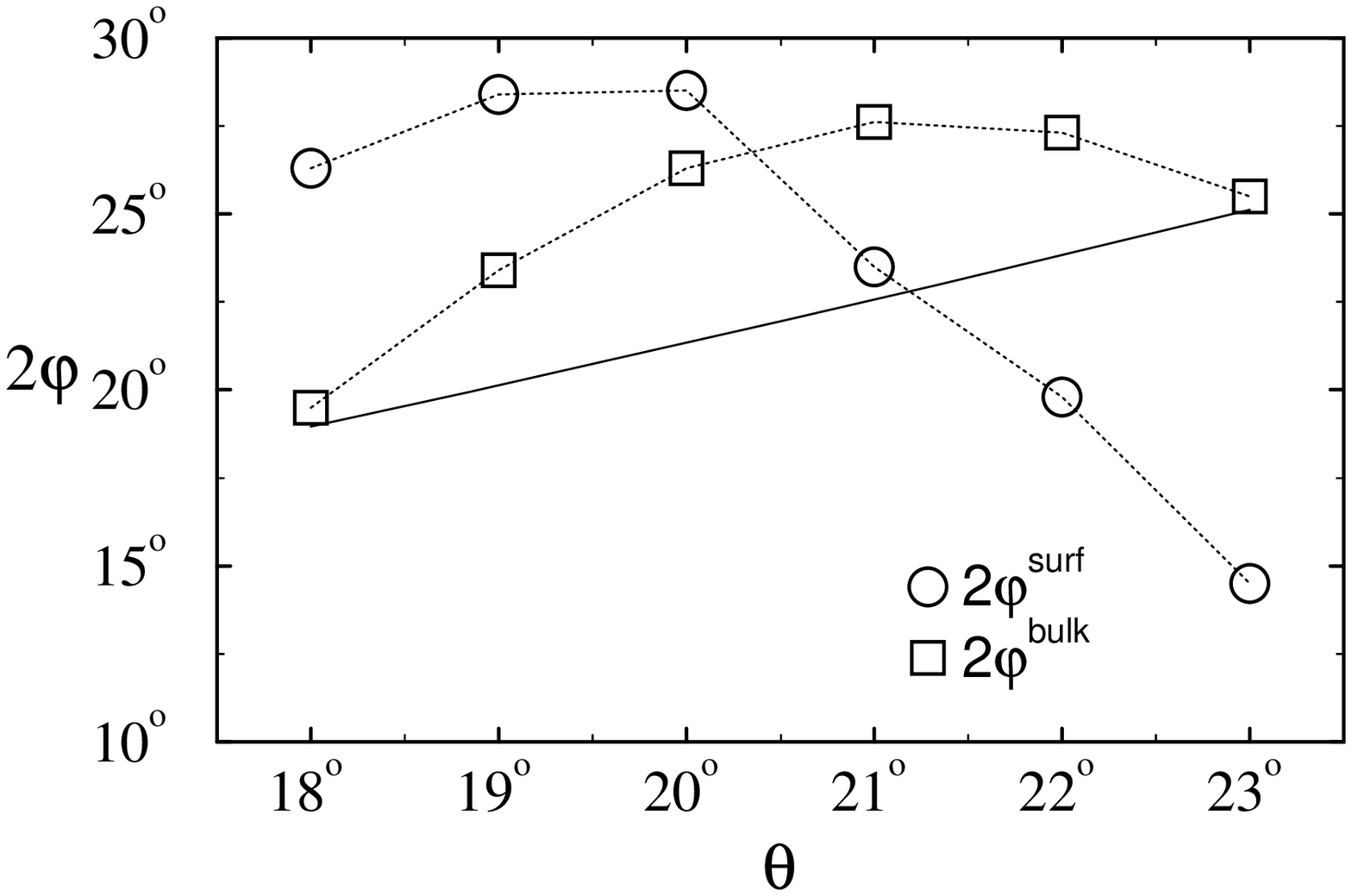}\hfil
\includegraphics[width=2.9in]{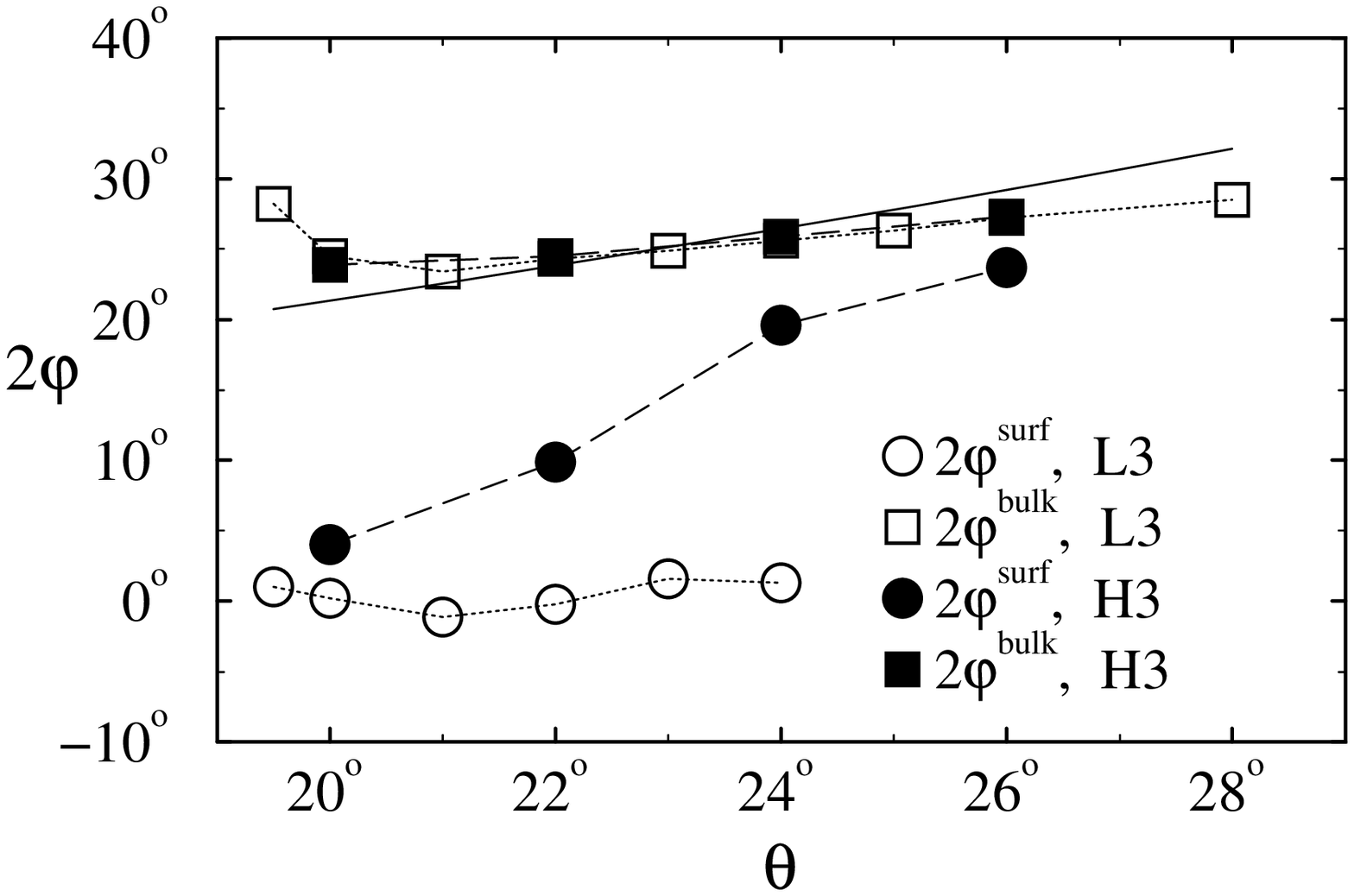}\\
\hfil(a)\hfil\hfil(b)\hfil\\
\bigskip
\caption{\em  The stress angle at the surface, $2\varphi^{\rm surf}$
  (circles), and in the bulk, $2\varphi^{\rm bulk}$ (squares), for
  (a) Model L2 (open symbols connected by dotted lines), 
  and (b) Model L3 (open symbols connected by dotted lines) and
  H3 (solid symbols connected by dashed lines). For Model L2, 
  the rheology at the surface $(z=h)$
  near $\theta_r$ is even farther away from the IYE condition 
  compared to the bulk. However, both 3D models observe near-IYE
  conditions at the surface near $\theta_r$, suggesting
  that the arrest of flow may be initiated by the surface rather 
  than the bulk. The solid lines depict behavior expected
  without a normal stress anomaly, when $\sin 2\varphi=\tan\theta$.}
\label{figMohr2}
\end{center} 
\end{figure}

In all cases, the behavior of $2\varphi$ as a function of depth 
can be fitted to an empirical form that starts at a ``surface'' 
value at the effective height $h$ and approaches a ``bulk'' 
value exponentially (see Fig.~\ref{figMohr1}b):
\begin{equation}
2\varphi(z)=2\varphi^{\rm bulk}+2(\varphi^{\rm surf}-
\varphi^{\rm bulk})e^{-(h-z)/\delta}.
\end{equation}
Figure~\ref{figMohr2} depicts the values for the fitting parameters
 $2\varphi^{\rm surf}$ and $2\varphi^{\rm bulk}$
as a function of tilt angle for the three main models 
studied in this paper.  The following observations can be made:

(i) $2\varphi$, and consequently all the ratios of stress tensor
components, becomes independent of depth below a transitional surface
layer about $5d$ to $8d$ in thickness. 

(ii) In 2D (Model L2, see Fig.~\ref{figMohr2}a), as $\theta$ is
lowered to $\theta_r$, the stress state at the surface moves even
farther from IYE compared to the bulk. Independent observations
confirm that the top surface does not play any discernible role in the
arrest and start of flow; this primarily occurs near the bottom
surface.

(iii) However, for both models in 3D (Fig.~\ref{figMohr2}b), 
as $\theta$ is 
lowered to $\theta_r$, the surface layer does approach incipient yield
($2\varphi=0$) while the bulk remains far from it.
It appears that the stabilization of the surface layer 
at $\theta=\theta_r$ is responsible for the arrest and subsequent
restart of flow in the entire system, accompanied by a near-elimination 
of flow hysteresis.

\end{multicols}

\begin{multicols}{2}

(iv) The bulk has
nearly identical normal stresses $\sigma_{xx}$ and $\sigma_{zz}$, 
which would have corresponded to $2\varphi=\arcsin(\tan\theta)$, 
depicted by the solid lines in Fig.~\ref{figMohr2}. In other words, 
the normal stress anomalies discussed in Sec.~\ref{secrheo} are quite 
small compared to what one would have attributed to a plastic material 
at incipient yield.  

(v) The transitional surface layer is not directly related to
the dilated layer; the former is much thicker near $\theta=\theta_r$ and
penetrates well into the region of constant density, as can be seen by
comparing Figs.~\ref{denandvel1} and \ref{figMohr1}b. In fact, 
upon approaching $\theta_r$, the width of the surface rheological layer
$\delta$ increases slightly whereas the width of the dilated 
layer decreases substantially. 

\begin{figure}
\begin{center}
\includegraphics[width=2.9in]{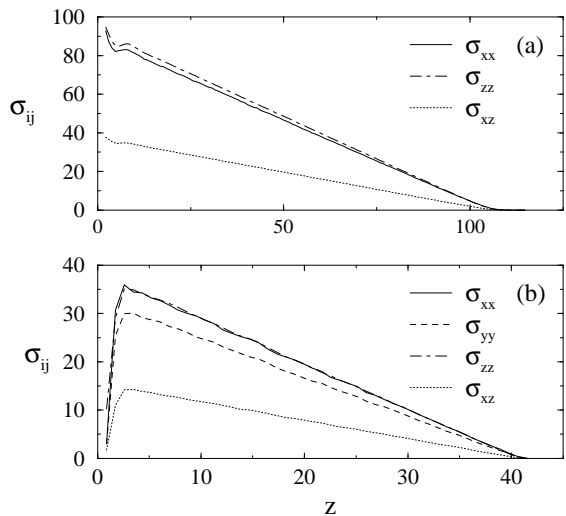}
\caption{\em Profiling the components of the stress tensor at 
$\theta=22^{o}$ in (a) Model L2 for $H=100$,
(b) Model L3 for $H=40$.} 
\label{stress1}
\end{center} 
\end{figure}

\subsection{Bulk Rheology}
\label{secrheo}

Having identified the behavior associated with the free surface at the
top, we can now investigate the stress tensor below this surface
layer. For tilt angles sufficiently above $\theta_r$, where the granular
medium behaves roughly like a fluid, one might expect the normal
stresses ($\sigma_{xx}, \sigma_{yy}, \sigma_{zz}$ in 3D,
$\sigma_{xx}~and~\sigma_{zz}$ in 2D) to be equal. In 3D, we find that
$\sigma_{yy}$ is smaller than the $xx-$ and $zz-$components by
$15-20\%$, suggesting that consolidation and compaction normal to the
shear plane is poorer.  The normal stresses and the driving shear
stress $\sigma_{xz}$, for the 2D and 3D linear spring model are shown
in Fig.~\ref{stress1}. The components ($\sigma_{xy}, \sigma_{yz}$) are
not shown, since they vanish due to the symmetries in the geometry;
they are indeed measured to be zero within the error bars associated
with the sample size and averaging time.

\begin{figure}
\begin{center}
\includegraphics[width=2.9in]{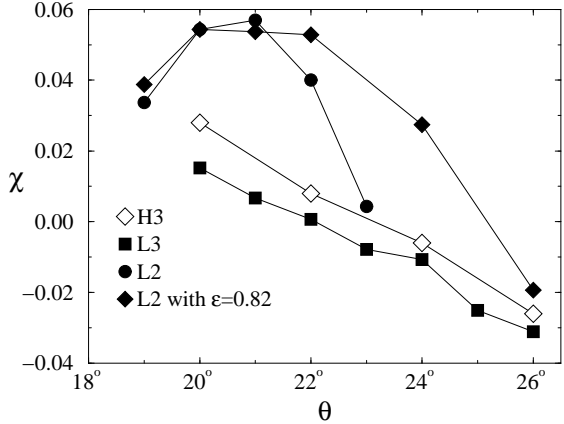}
\caption{\em Dependence of the normal stress anomaly 
$\chi$ on tilt angle $\theta$ for Models L2 and L3 
(closed symbols), and Model H3 (open symbols).}
\label{stress4} 
\end{center} 
\end{figure}

We observe that for both the 2D and 3D systems (and for both linear
spring and Hertz models), although $\sigma_{xx}\approx\sigma_{zz}$,
there are small but systematic deviations from perfect equality
that become independent of depth in the bulk.  
Let us define a ``normal stress
anomaly'' $\chi$ as,
\begin{equation}
\chi\equiv\frac{\sigma_{zz}-\sigma_{xx}}{\sigma_{zz}}.
\end{equation}
This is simply an alternate parameterization of the stress angle
$2\varphi$ defined earlier, introduced as a convenience to 
emphasize the small deviations around fluid-like behavior, for 
which $\chi=0$. Therefore, $\chi$ is also independent of height $z$ 
except near the top and bottom surfaces.  
With this in mind, we plot the bulk 
value of $\chi$ vs. $\theta$ in Fig.~\ref{stress4}, noting a strong 
angle dependence, in which $\chi$ is neither monotonic in $\theta$
nor of a specific sign. We have evaluated a class of 
homogeneous, polynomial, rotationally invariant constitutive 
stress-strain rate relations, but have not been able to 
satisfactorily describe these rather peculiar normal stress anomalies.

\begin{figure}
\begin{center}
\includegraphics[width=2.9in]{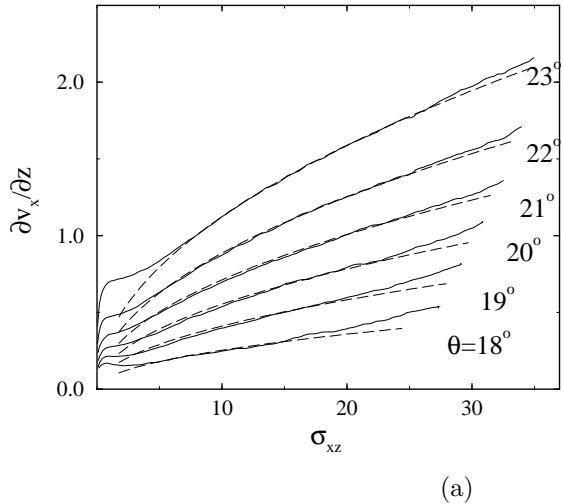}\hfil
\includegraphics[width=2.9in]{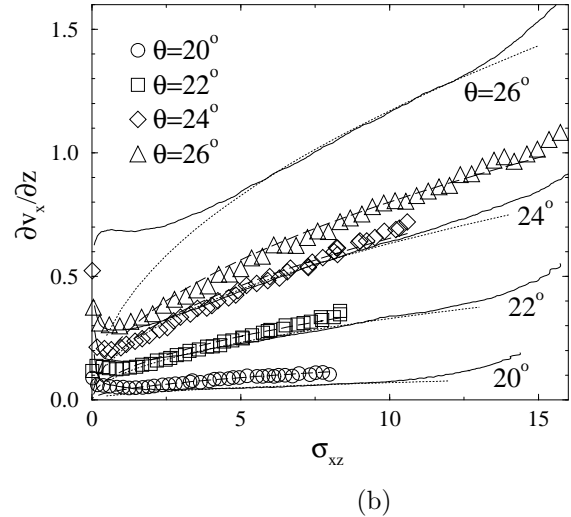}\\
\hfil(a)\hfil\hfil(b)\hfil\\
\bigskip
\caption{\em Rheology curves of chute flow systems-shear strain  
  vs. shear stress; a) 2D, (Model L2, $H=100$) and b) 3D ($H=40$):
  Model L3 (symbols) and Model H3 (solid lines).  Bagnold scaling fits
  in the bulk are shown by dashed lines for Model L and dotted lines
  for Model H.  }
\label{stress5}  
\end{center}  
\end{figure}

\begin{figure}
\begin{center}
\includegraphics[width=2.9in]{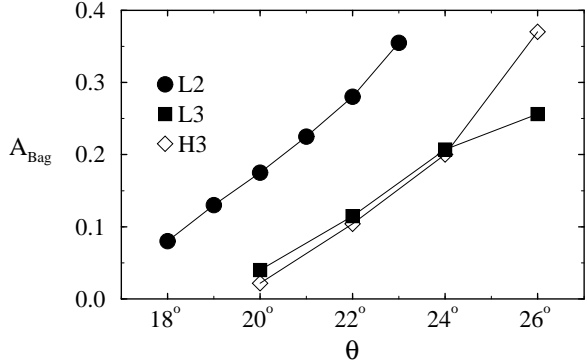}
\caption{\em Strain rate amplitude 
$A_{\rm Bag}=\dot\gamma/\sqrt{\sigma_{xz}}$ 
associated with the bulk Bagnold rheology for the three model systems.
Whereas 3D amplitudes extrapolate to zero at $\theta_r$, there is a
finite jump associated with the 2D amplitude at $\theta_r$.}
\label{bagnoldamp}  
\end{center}  
\end{figure}

The fact that the stress varies linearly with depth and our earlier
observations of constant density suggests that the analysis relevant
to our systems is that due to Bagnold \cite{bagnold1}. Bagnold's
collisional-momentum transfer analysis for granular systems works
under the assumption of a constant density profile, resulting in
stress profiles that vary linearly with depth. The essence of
Bagnold's theory is a constitutive equation whereby the shear stress
$\sigma_{xz}$ is proportional to the square of the strain rate
$\dot{\gamma}^{2}\equiv\left (\partial v_{x}(z)/\partial z\right
)^{2}$, where $v_{x}(z)$ is the velocity in the direction of flow 
at height $z$:
\begin{equation}
\sigma_{xz}=A_{\rm Bag}^2 {\dot\gamma}^2.
\end{equation}
Combined with Eq.(\ref{eqsxz}), and the no-slip boundary condition at
$z=0$, this results in a velocity profile of the form,
\begin{equation}
v_{x}(z)=A_{\rm Bag}h^{3/2}\left(\frac{2}{3}\sqrt{\rho g \sin\theta}
\right)\left[1-\left(\frac{h-z}{h}\right)^{3/2}\right].
\end{equation}
\noindent
From Fig.~\ref{stress5}, we observe that for
the bulk of the flow, the relationship
$\sigma_{xz}\propto\dot{\gamma}^{2}$ holds to a good approximation
below the first 5-8 layers, and away from the bottom wall, for the 2D
and 3D systems. We have fitted the ``Bagnold'' scaling with the dotted
lines, the solid lines and symbols represent the simulation data
\cite{footnote3}. 
The tilt dependence of the overall amplitude of the strain rate, 
$A_{\rm Bag}$, is shown in Fig.~\ref{bagnoldamp}. In 3D, $A_{\rm Bag}$
continuously approaches zero at the angle of repose, whereas in 2D, 
there is a jump in this amplitude, consistent with the overall hysteretic
behavior. 

\end{multicols}

\vspace{1in}

\begin{multicols}{2}

\begin{figure}
\begin{center}
\includegraphics[width=2.9in]{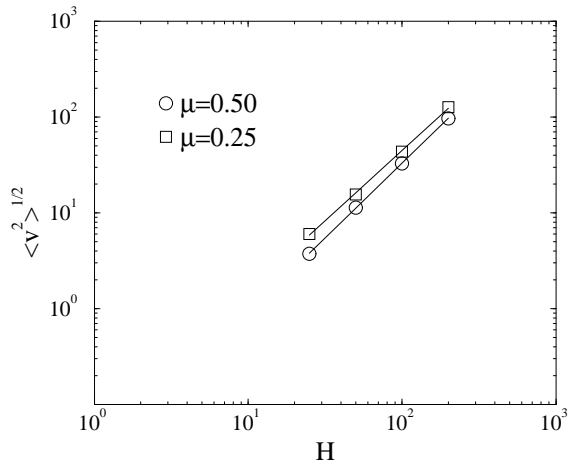}\hfil
\includegraphics[width=2.9in]{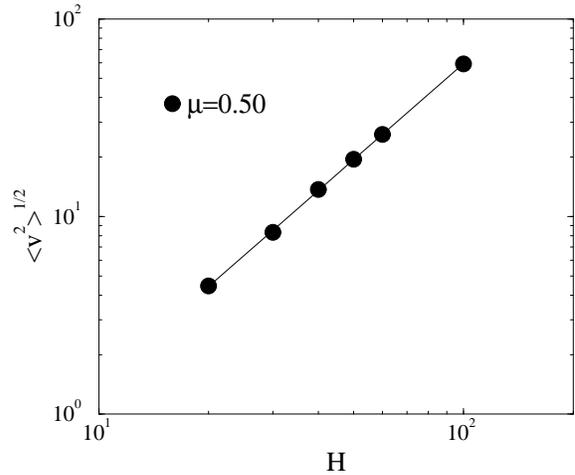}\\
\hfil(a)\hfil\hfil(b)\hfil\\
\bigskip
\caption{\em Scaling of velocity in the direction of flow 
  $<v^{2}_{x}(z)>^{1/2}$ with system height $H$ in; (a) 2D at $20^{o}$
  with $\epsilon=0.92$, for two different values of $\mu$, (b) model
  H3 at $24^{o}$. The slope of the lines indicate that
  $<v^{2}>^{1/2}\propto H^{\alpha}$, with $\alpha=1.52\pm0.05$.}
\label{denandvel22} 
\end{center} 
\end{figure}

Another way to test this scaling is by plotting the average 
velocity $<v^{2}>^{1/2}$ as a function of $H$
(which is proportional to $h$). The scaling in Fig.~\ref{denandvel22}
shows that $<v^{2}>^{1/2}\propto H^{\alpha}$, where $\alpha=1.52\pm
0.05$. This result also agrees well with experiment
\cite{pouliquen1}.  If we rescale the data from Fig.~\ref{denandvel2},
we find good agreement apart from the region near the top surface
where the density is no longer constant. This suggests that Bagnold's
theory may provide an approximate description of the bulk motion of
our systems.  In fact, Bagnold scaling is a generic dimensional result
for the situation where the time scale of the system is only set by
the inverse of the shear rate, as is the case here \cite{deniz2}.

Bagnold's original stress-strain rate relationship arises from a 
momentum transfer mechanism that is based on binary collisions. 
From the simulation data, we
find that the dominant term in the stress is due to lasting 
contacts between particles, and the ballistic (kinetic) 
contribution to the stress is
significantly smaller (about $1\%$ of the total value).
Thus, the success of the Bagnold scaling is based on the dimensional 
structure of the problem, rather than on the particular
momentum transfer mechanism that he identified. 
\end{multicols}

\begin{figure}
\begin{center}
  \includegraphics[width=2.9in]{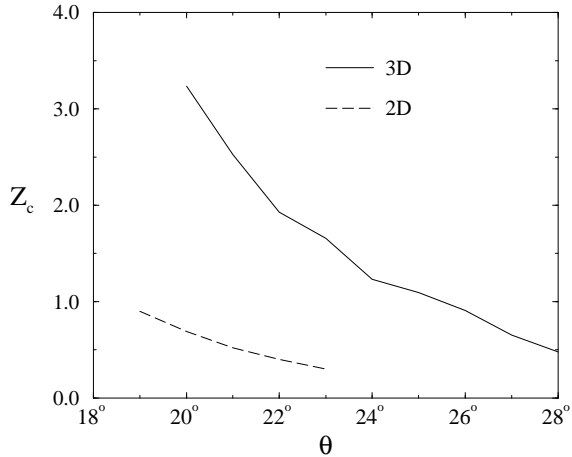}
\caption{\em Averaged instantaneous coordination number $Z_{c}$,as 
a function of tilt angle for Model L2 with $H=100$, and 
Model L3 with $H=40$.} 
\label{contactang} 
\end{center}
\end{figure}

\begin{multicols}{2}
Another method to test the nature of collisions is to 
compute the average coordination number $Z_{c}$ as a function of
inclination angle. This data, for the 2D and 3D linear-spring systems,
is shown in Fig.~\ref{contactang}. In a system dominated by binary 
collisions, one would expect $Z_{c}\ll 1$; this is clearly not the 
case for our system. The observed behavior  
is an increasing $Z_{c}$ as $\theta$ approaches the angle of repose
from above. 
Normalized this way for a static 2D triangular lattice with no
free particles, the value would be 3. Similarly, for 3D 
static packings one
might expect a value between $4-6$ \cite{Makse}. 

Because of these observations, we
reason that contributions to the kinetic term of the stress tensor do
not play a significant role in determining the macroscopic quantities
measured. It might then be argued that for a densely packed pile of 
stiff objects in motion, the overall time evolution of the system in 
the configurational phase space is primarily constrained and
controlled by aspects of geometrical packing,
rather than the specific form of the stiff force laws between 
particles or dissipation functions. This might be why
the system is so insensitive to variations 
in the interaction parameters, as described in Sec.~\ref{intparam}.

\section{Conclusions} 
\label{secconc}

We have concentrated on the steady-state nature of chute flows,
specifying first the region in phase space in which such flows can be
observed, and second the structure and rheology of these flows.
 
A region of constant packing fraction is a generic feature in our 2D
and (two) 3D models, with only a small dilated layer at the free
surface.  Analysis of the velocity profiles has revealed that to a
good approximation, Bagnold scaling holds: the Bagnold velocity 
profile, $v_{x}\propto H^{1.5}$, and rheology,
$\sigma\propto\dot{\gamma}^2$, is reasonably
verified away from the surface.  This is in contrast with earlier
simulations on chute flows, which indicated linear velocity profiles.
We argue that although this latter may be the case for small systems, 
such as flowing layers less than 20 particles high, steady flows of 
moderately thick systems are well-approximated by Bagnold scaling.

Although the regime of Bagnold-like flow appears to dominate the
system, we have found that deviations from this simple theory exist.
The normal stress anomaly remains a mystery, and our fits to the
stress-strain rate curves apply only away from the top and bottom
surfaces. We have also found that the transmission of stress in such
dense flows is dominated by contacts, as opposed to binary collisions
in Bagnold's analysis of dilute flows.

Finally, we observe that the normal stresses in bulk flows do not
approach a Coulomb yield criterion structure at the angle of repose,
despite the continuous disappearance of the shear rate at this
threshold. The fact that Coulomb yield is approached at the surface
for 3D flows hints at a special role for surface failure in this case.

Our simulation code, both in its simple and parallelised versions,
enables us to study large systems for very long time scales, and we
continue to investigate some of the outstanding issues in this area.
We will report elsewhere the differences between rough and smooth
bottom surfaces \cite{leo8}.  We will also go on to study 3D planar
Couette flows, extending Ref.~\cite{grest1}, and will be reporting on
this in the future.

\section*{Acknowledgements}
DL is supported by the Israel Science Foundation under grant 211/97.
Sandia is a multiprogram laboratory operated by Sandia Corporation, a
Lockheed Martin Company, for the United States Department of Energy
under Contract DE-AC04-94AL85000.

\end{multicols}

\begin{thebibliography}{10}
\bibitem{footnote1}
Although we do not see slip in this study, we have observed slip in our 2D
  simulations using a smoother bottom wall constructed from spheres of the same
  diameter as those in the bulk.
\bibitem{footnote3}
Although the overall fit is good, deviations from the simple scaling law occur
  even at depths where the density has reached its bulk value. Thus, the
  rheology of the system does not appear to be entirely local, with surface
  effects penetrating farther into the pile than is suggested by the density
  profile.
\end{thebibliography}
\end{document}